\documentclass[draftclsnofoot,onecolumn,12pt]{IEEEtran}
\usepackage[nospace,noadjust]{cite}

\usepackage{amssymb,amsmath}
\usepackage{graphics}
\usepackage{float}
\usepackage{graphicx}
\usepackage{amsmath}
\usepackage{cite}
\usepackage{color}
\usepackage{dsfont}
\usepackage{bm}
\usepackage{upgreek}
\usepackage{arydshln}
\usepackage{enumerate}
\usepackage{amsthm}
\usepackage{graphicx}
\usepackage{threeparttable}
\usepackage{caption}
\usepackage{multirow}
\usepackage{color}
\usepackage{xfrac}
\usepackage{array}
\usepackage{nomencl}
\usepackage{hyphenat}

\newcommand{\be}{\begin{equation}}
\newcommand{\ee}{\end{equation}}
\newcommand{\innerprod}[2]{ \left\langle #1 , #2\right\rangle}

\newcommand{\norm}[1]{ || #1 ||}
\newcommand{\mb}[1]{\mathbf{#1}}
\newcommand{\bs}[1]{\boldsymbol{#1}}
\newcommand{\lbr}{\left\lbrace}
\newcommand{\rbr}{\right\rbrace}
\newcommand{\virg}[1]{\textquotedblleft#1\textquotedblright}
\newcommand{\vsigma}{\mathrm{vecs}(\bs{\Sigma})}
\newcommand{\vsigmat}{\mathrm{vecs}(\bs{\Sigma}_0)}

\newcommand{\vcsigmatinv}{\mathrm{vec}(\bs{\Sigma}_0^{-1})}
\newcommand{\kronsigmatinv}{\bs{\Sigma}_0^{-1}\otimes\bs{\Sigma}_0^{-1}}
\newcommand{\kronsigmatinvm}{\bs{\Sigma}_0^{-1/2}\otimes\bs{\Sigma}_0^{-1/2}}

\newtheorem{theorem}{Theorem}[section]

\newtheorem{definition}{Definition}[section]
\newtheorem{proposition}{Proposition}[section]

\begin{document}
%
\title{Semiparametric Inference and Lower Bounds for Real Elliptically Symmetric Distributions}
%
%
%

\author{Stefano~Fortunati,~\IEEEmembership{Member,~IEEE,}
        Fulvio~Gini,~\IEEEmembership{Fellow,~IEEE,}
        Maria~S.~Greco,~\IEEEmembership{Fellow,~IEEE,}
        Abdelhak~M.~Zoubir,~\IEEEmembership{Fellow,~IEEE}
        and~Muralidhar~Rangaswamy,~\IEEEmembership{Fellow,~IEEE} 
\thanks{S. Fortunati is with Universit\`a di Pisa, Dipartimento di Ingegneria dell'Informazione, Pisa, Italy and with Technische Universit\"at Darmstadt, Signal Processing Group, Darmstadt, Germany 
(e-mail: stefano.fortunati@iet.unipi.it).}
\thanks{F. Gini and M. S. Greco are with Universit\`a di Pisa, Dipartimento di Ingegneria dell'Informazione, Pisa, Italy (e-mail: f.gini,m.greco@iet.unipi.it).}
\thanks{A. M. Zoubir is with Technische Universit\"at Darmstadt, Signal Processing Group, Darmstadt, Germany (e-mail: zoubir@spg.tu-darmstadt.de).}
\thanks{M. Rangaswamy is with U.S. AFRL, Sensors Directorate, Wright-Patterson AFB, OH, USA (e-mail: muralidhar.rangaswamy@us.af.mil).}}

%



\maketitle

\begin{abstract}
	
This paper has a twofold goal. The first aim is to provide a deeper understanding of the family of the Real Elliptically Symmetric (RES) distributions by investigating their intrinsic semiparametric nature. The second aim is to derive a semiparametric lower bound for the estimation of the parametric component of the model. The RES distributions represent a semiparametric model where the parametric part is given by the mean vector and by the scatter matrix while the non-parametric, infinite-dimensional, part is represented by the density generator. Since, in practical applications, we are often interested only in the estimation of the parametric component, the density generator can be considered as nuisance. The first part of the paper is dedicated to conveniently place the RES distributions in the framework of the semiparametric group models. The second part of the paper, building on the mathematical tools previously introduced, the Constrained Semiparametric Cram\'{e}r-Rao Bound (CSCRB) for the estimation of the mean vector and of the constrained scatter matrix of a RES distributed random vector is introduced. The CSCRB provides a lower bound on the Mean Squared Error (MSE) of any robust $M$-estimator of mean vector and scatter matrix when no a-priori information on the density generator is available. A closed form expression for the CSCRB is derived. Finally, in simulations, we assess the statistical efficiency of the Tyler's and Huber's scatter matrix $M$-estimators with respect to the CSCRB.
\end{abstract}


\section{Introduction}
A prerequisite for any statistical inference method is the notion of a \textit{statistical model}, say $\mathcal{P}$, i.e. a collection, or a family, of probability density functions (pdfs) that is able to characterize random phenomena based on their observations. The most widely used models are \textit{parametric models}. A parametric model is a family of pdfs parametrized by the elements of a subset $\Gamma$ of a finite-dimensional Euclidean space $\mathbb{R}^q$. The popularity of parametric models is due to the ease of derivation of inference algorithms. On the other hand, a major drawback is their \virg{narrowness} that can lead to misspecification problems \cite{SPM}, \cite{ISI2017}. Their counterpart are \textit{nonparametric models}, a wide family of pdfs that can be required to satisfy some functional constraints, e.g. symmetry, smoothness or moment constraints. While the use of a nonparametric model minimizes the risk of model misspecification, the amount of data needed for nonparametric inference may represent an insurmountable obstacle in practical applications. \textit{Semiparametric models} have been introduced as a compromise between the \virg{narrowness} of parametric models and the cost of using nonparametric ones \cite{Bickel_Intro}. More formally, let $\mb{x}$ be a random vector taking values in the sample space $\mathcal{X} \subseteq \mathbb{R}^N$. Then, a semiparametric model $\mathcal{P}_{\bs{\gamma},l}$ is a family of pdfs parametrized by a finite-dimensional parameter vector of interest $\bs{\gamma} \in \Gamma \subseteq \mathbb{R}^q$, along with an infinite-dimensional nuisance parameter $l \in \mathcal{L}$, where $\mathcal{L}$ is a set of functions:
\be
\mathcal{P}_{\bs{\gamma},l} \triangleq \lbr p_X | p_X(\mb{x}|\bs{\gamma},l), \bs{\gamma} \in \Gamma, l \in \mathcal{L}\rbr.
\ee     

There is a rich statistical literature on semiparametric models and their applications. For a comprehensive and detailed list of the main contributions in this field, we refer the reader to \cite{Bickel_Intro} and \cite{Wellner85} and to the seminal book \cite{BKRW}. However, this profound theoretical understanding of semiparametric models has not been fully exploited in Signal Processing (SP) problems as yet. Two, among the very few, examples of SP applications of the semiparametric inference are the references \cite{Amari} and \cite{Zoubir_semi}, where the semiparametric theory has been applied to blind source separation and nonlinear regression, respectively. 

This paper aims at improving the understanding of potential applications of semiparametric models. Specifically, we focus our attention on the joint estimation of the mean vector $\bs{\mu}$ and of the (constrained) scatter matrix $\bs{\Sigma}$ in the family of Real Elliptically Symmetric (RES) distributions by providing a closed form expression, up to a (numerically performed) singular value decomposition, for the Constrained Semiparametric Cram\'{e}r-Rao Bound (CSCRB) on the MSE of any robust estimator of $\bs{\mu}$ and $\bs{\Sigma}$. As we will discuss below, a constraint on Sigma is required to avoid the scale ambiguity that characterizes the definition of scatter matrix in RES distributions. The RES class represents a wide family of distributions that includes the Gaussian, the $t$, the Generalized Gaussian and all the real Compound-Gaussian distributions as special cases (\cite{CAMBANIS1981,RichPhD,RES_rev,Yao,zozor,kotz,Esa}, and \cite[Ch.~4]{book_zoubir}). The elliptical distributions are of fundamental importance in many practical applications since they can be successfully exploited to statistically characterize the non-Gaussian behavior of noisy data. Moreover, as we will discuss below, RES distributions represent an example of a semiparametric model where the parametric part is represented by the mean vector and by the scatter matrix. They should be estimated in the presence of an infinite-dimensional nuisance parameter, i.e. the density generator, which is generally unknown.

This paper is the natural follow on of our previous work \cite{For_EUSIPCO}. In \cite{For_EUSIPCO}, we provided, in a tutorial and accessible manner, a general introduction to semiparametric inference framework and to the underling mathematical tools needed for its development. Moreover, the extension of the classical CRB in the presence of a finite-dimensional nuisance deterministic vector to the SCRB where the nuisance parameter belongs to a certain, infinite-dimensional, function space has been reported in Theorem 1 in \cite{For_EUSIPCO}. As discussed in the statistical literature and summarized in \cite{For_EUSIPCO}, this generalization can be carried out by means of three key elements:
\begin{itemize}
	\item the Hilbert space $\mathcal{H}^q$ of all the $q$-dimensional, zero-mean, vector-valued function of the data vector,
	\item a notion of \textit{tangent space} $\mathcal{T}$ for a statistical model,
	\item an orthogonal projection operator on $\mathcal{T}$, i.e. $\Pi(\cdot|\mathcal{T})$. 
\end{itemize}  
A formal definition of the Hilbert space $\mathcal{H}^q$ and of the projection operator $\Pi(\cdot|\mathcal{T})$ can be found in Appendix \ref{Hilber_random}, while the tangent space for both parametric and semiparametric models has been defined, in a tutorial manner, in \cite{For_EUSIPCO}. 
      
This paper aims at investigating the applications of the general concepts introduced in \cite{For_EUSIPCO} to the semiparametric model generated by RES distributions. We start by introducing the semiparametric group model and then we continue by showing that the RES class actually possesses this structure. Building on the mathematical framework that characterizes semiparametric group models and, in particular, their tangent space and projection operator, we then show how to derive the CSCRB for the joint estimation of the mean vector $\bs{\mu}$ and the scatter matrix $\bs{\Sigma}$ of a RES distributed random vector.  

The problem of establishing a semiparametric lower bound for the joint estimation of $\bs{\mu}$ and $\bs{\Sigma}$ in the RES class has been investigated firstly by Bickel in \cite{Bickel_paper}, where a bound on the estimation error of the inverse of the scatter matrix has been derived. More discussions and analyses have also been presented in \cite{BKRW} (Sec. 4.2 and Sec. 6.3). More recently, in a series of papers (\cite{Hallin_P_Annals}, \cite{Hallin_Annals_Stat_2}, \cite{Hallin_P_2006} and \cite{PAINDAVEINE}), Hallin and Paindaveine rediscovered the RES class as a semiparametric model and, by using two approaches based on Le Cam's theory \cite{LeCam} and on rank-based invariance \cite{Hallin_Werker}, they presented the SCRB for the joint estimation of $\bs{\mu}$ and $\bs{\Sigma}$ in its most general form. However, even if valuable and profound, Hallin and Paindavaine's work requires a deep understanding of Le Cam's theory on Local Asymptotic Normality \cite{LeCam}. For this reason, starting from the results in \cite{BKRW} (Sec. 4.2 and Sec. 6.3), we propose here an alternative derivation of the SCRB by using a simpler, even if less general, approach.

Along with the derivation of a lower bound on the estimation performance, we always have to specify the class of estimators to which such bound applies. It can be shown that the SCRB is a lower bound to the MSE of any \textit{regular} and \textit{asymptotic linear} (RAL) estimator (see \cite[Sec. 2.2]{BKRW}, \cite{Bickel_paper}, \cite{Newey}, \cite{newey_paper}, \cite{klaassen1987}, \cite[Ch. 3]{Tsiatis} and \cite[Ch. 4]{Rieder} for additional details). Even if we do not address this issue here, it must be highlighted that the class of RAL estimators is a very wide family that encompasses the Maximum Likelihood estimator and all the $R$-, $S$-, and in particular, $M$- robust estimators.

The rest of the paper is organized as follows. In Sec. \ref{semi_group_mod} the semiparametric group model is presented with a particular focus on the calculation of the tangent space and projection operator. Sec. \ref{RES_section} collects the basic notions on RES distributions and their intrinsic semiparametric-group structure is investigated. The step-by-step derivation of the CSCRB for the estimation of $\bs{\mu}$ and $\bs{\Sigma}$ is provided in Sec. \ref{SCRB_for_RES}. The efficiency of the Sample Covariance Matrix and of two robust scatter matrix $M$-estimators, Tyler's and the Huber's estimators, is assessed in Sec. \ref{simulation_RES} using the previously derived CSCRB. Finally, some concluding remarks are collected in Sec. \ref{conclusions}.

\textit{Notation}: Throughout this paper, italics indicates scalars or scalar-valued functions ($a,A$), lower case and upper case boldface indicate column vectors ($\mb{a}$) and matrices ($\mb{A}$) respectively. Note that, since we deal with Hilbert spaces, the word \virg{vector} indicates both Euclidean vectors and vector-valued functions. For clarity, we indicate sometimes a vector-valued function as $\mb{a}\equiv\mb{a}(\mb{x})$. Each entry of a matrix $\mathbf{A}$ is indicated as $a_{i,j}\triangleq [\mathbf{A}]_{i,j}$. The superscript $T$ indicates the transpose operator. Finally, for random variables or vectors, the notation $=_d$ stands for "has the same distribution as". 

\section{The semiparametric group models}
\label{semi_group_mod}
This section introduces a particular semiparametric model, i.e. the \textit{semiparametric group model}. As the name suggests, this class of semiparametric models is generated by the action of a group of invertible transformations on a random vector whose pdf is allowed to vary in a given set. As we will show in the sequel, this group-based data generating process allows for an easy calculation of the nuisance tangent space and of the orthogonal projection operator. Before introducing the definition of this class of semiparametric models, let us first introduce some related notation. 

Let $\mathcal{A}$ be a group of invertible transformations from $\mathbb{R}^N$ into itself. Suppose that each transformation $\alpha \in \mathcal{A}$ can be parametrized by means of a real vector $\bs{\gamma} \in \Gamma \subseteq \mathbb{R}^q$, i.e.
\be
\label{par_group}
\mathcal{A} = \{\alpha|\alpha(\cdot;\bs{\gamma}) \triangleq \alpha_{\bs{\gamma}}(\cdot); \bs{\gamma} \in \Gamma\}.
\ee

We will indicate with $\alpha_{\bs{\gamma}}^{-1}$ the inverse of $\alpha_{\bs{\gamma}}$. The operation $\alpha_{\bs{\gamma}_2} \circ \alpha_{\bs{\gamma}_1}$ denotes the composition of $\alpha_{\bs{\gamma}_1}$ and $\alpha_{\bs{\gamma}_2}$ that can be explicitly expressed as $(\alpha_{\bs{\gamma}_2} \circ \alpha_{\bs{\gamma}_1})(\cdot) \triangleq \alpha_{\bs{\gamma}_2}(\alpha_{\bs{\gamma}_1}(\cdot))$. Finally, $\bs{\gamma}_e$ indicates the parameter vector that characterizes the identity transformation $\alpha_{\bs{\gamma}_e}$, such that $\alpha_{\bs{\gamma}_e}(\cdot)=\cdot$.

\begin{definition}
	\label{def_group_model}
	\textit{(see \cite[Sec. 4.2]{BKRW})} Let $\mb{z} \in \mathbb{R}^N$ be a real-valued random vector with pdf $p_Z$, i.e. $\mb{z} \sim p_Z(\mb{z})$. The parametric group model $\mathcal{P}_{\bs{\gamma}}$, generated by the action of the group of invertible parametric transformations $\mathcal{A}$, given in \eqref{par_group}, on the random vector $\mb{z} \sim p_Z(\mb{z})$, is the set of parametric pdfs of the transformed random vector $\alpha_{\bs{\gamma}}(\mb{z}) = \mb{x} \sim  p_X(\mb{x}|\bs{\gamma})$. Specifically, $\mathcal{P}_{\bs{\gamma}}$ can be explicitly expressed as:
	\be
	\mathcal{P}_{\bs{\gamma}} = \lbr p_X | p_X(\mb{x}|\bs{\gamma}) = |\mb{J}(\alpha^{-1}_{\bs{\gamma}})(\mb{x})| p_Z(\alpha^{-1}_{\bs{\gamma}}(\mb{x}));   \bs{\gamma} \in \Gamma \rbr,
	\ee
	where $[\mb{J}(\alpha^{-1}_{\bs{\gamma}})(\mb{x})]_{i,j} \triangleq \partial [\alpha^{-1}(\mb{x};\bs{\gamma})]_i/\partial \gamma_j$ is the Jacobian matrix of the inverse transformation $\alpha^{-1}_{\bs{\gamma}}$ and $|\cdot|$ defines the absolute value of the determinant of a matrix. The generalization to semiparametric models can be obtained by allowing the pdf $p_Z$ to vary within a large set of density functions $\mathcal{L}$. Consequently, a semiparametric group model generated by the parametric group $\mathcal{A}$ in \eqref{par_group} can be expressed as:
	\be
	\label{semi_group_model}
	\mathcal{P}_{\bs{\gamma},p_Z} = \lbr p_X | p_X(\mb{x}|\bs{\gamma},p_Z) = |\mb{J}(\alpha^{-1}_{\bs{\gamma}})(\mb{x})| p_Z(\alpha^{-1}_{\bs{\gamma}}(\mb{x}));  \bs{\gamma} \in \Gamma, p_Z \in \mathcal{L} \rbr.
	\ee
\end{definition}
Using the notation introduced in \cite{For_EUSIPCO}, the \textit{actual} \virg{semiparametric vector} is indicated as $(\bs{\gamma}_0^T,p_{Z,0})^T$ and, consequently, the true pdf is given by:
\be
\label{true_pdf_semipar_group}
p_0(\mb{x}) \triangleq p_X(\mb{x}|\bs{\gamma}_0,p_{Z,0})= |\mb{J}(\alpha_{\bs{\gamma}_0}^{-1})(\mb{x})| p_{Z,0}(\alpha_{\bs{\gamma}_0}^{-1}(\mb{x})).
\ee
Moreover, from now on, we denote by $E_0\{\cdot\}$ the expectation operator with respect to the \textit{true} pdf $p_0(\mb{x})$. Note that the word \virg{true} here indicates the actual pdf, and consequently the actual semiparametric vector, that characterizes the data.

As mentioned before, the most useful feature of a semiparametric group model is the fact that its underling group structure allows for a convenient derivation of the nuisance tangent space and of the relevant projection operator. The following proposition formalizes this concept.

Let $\mathcal{P}_{\bs{\gamma},p_Z}$ be a semiparametric group model defined in \eqref{semi_group_model}. Let $\mathcal{T}_{p_{Z,0}}(\bs{\gamma})$ be the semiparametric nuisance tangent space of $\mathcal{P}_{\bs{\gamma},p_Z}$ evaluated at $(\bs{\gamma}^T,p_{Z,0})^T$, where $\bs{\gamma}$ is a generic element of the finite-dimensional parameter space $\Gamma$. Let us indicate by $\mathcal{T}_{p_{Z,0}}(\bs{\gamma}_e) $ the semiparametric nuisance tangent space of $\mathcal{P}_{\bs{\gamma},p_Z}$ evaluated at $(\bs{\gamma}_e^T,p_{Z,0})^T$, where, as defined before, $\bs{\gamma}_e$ is the parameter vector that characterizes the identity transformation.
\begin{proposition}
	\label{semi_group_tan_proj}
	Let $\mb{t}$ be a generic $q$-dimensional vector-valued function belonging to $\mathcal{T}_{p_{Z,0}}(\bs{\gamma}_e)$, then the semiparametric nuisance tangent space $\mathcal{T}_{p_{Z,0}}(\bs{\gamma})$ can be obtained from $\mathcal{T}_{p_{Z,0}}(\bs{\gamma}_e)$ as follows:
	\be
	\label{semipar_group_tan}
	\mathcal{T}_{p_{Z,0}}(\bs{\gamma}) = \lbr \mb{t} \circ \alpha_{\bs{\gamma}}^{-1} | \mb{t} \in \mathcal{T}_{p_{Z,0}}(\bs{\gamma}_e) \rbr, \forall \bs{\gamma} \in \Gamma.
	\ee
	Moreover, let $\mb{l}$ be a generic $q$-dimensional vector-valued function in $\mathcal{H}^q$, then the projection operator $\Pi(\cdot|\mathcal{T}_{p_{Z,0}}(\bs{\gamma}))$ on $\mathcal{T}_{p_{Z,0}}(\bs{\gamma})$ (see Appendix \ref{Hilber_random}) can be obtained form the projection operator $\Pi(\cdot|\mathcal{T}_{p_{Z,0}}(\bs{\gamma}_e))$ on $\mathcal{T}_{p_{Z,0}}(\bs{\gamma}_e)$ as follows:
	\be
	\label{semipar_group_proj}
	\Pi(\mb{l}|\mathcal{T}_{p_{Z,0}}(\bs{\gamma})) = \Pi(\mb{l} \circ \alpha_{\bs{\gamma}}|\mathcal{T}_{p_{Z,0}}(\bs{\gamma}_e)) \circ \alpha_{\bs{\gamma}}^{-1}, \forall \bs{\gamma} \in \Gamma.
	\ee
\end{proposition} 
The proof can be found in \cite[Sec. 4.2, Lemma 3]{BKRW}.

It is worth noticing that Proposition \ref{semi_group_tan_proj} can be directly used to derive the nuisance tangent space at the true semiparametric vector $(\bs{\gamma}_0^T,p_{Z,0})^T$, i.e. $\mathcal{T}_{p_{Z,0}} \equiv \mathcal{T}_{p_{Z,0}}(\bs{\gamma}_0)$ and the relevant projection operator $\Pi(\cdot|\mathcal{T}_{p_{Z,0}})$. This can be done by evaluating the relations \eqref{semipar_group_tan} and \eqref{semipar_group_proj} at the true parameter vector of interest $\bs{\gamma}_0$. As discussed below, Proposition \ref{semi_group_tan_proj} is of fundamental importance for the derivation of the Semiparametric Cram\'{e}r-Rao Bound (SCRB) for $\bs{\gamma}_0$ in the class of RES distributions.

\section{The family of RES distributions as a semiparametric model}
\label{RES_section}

In this section, the semiparametric nature of the family of RES distributions is investigated. In particular, we show that the RES class can be conveniently interpreted as a semiparametric group model. Here we restrict the discussion to the \textit{absolutely continuous case} \cite[Sec. III.D]{Esa}, i.e. we always assume that a RES distributed random vector admits a pdf. In what follows, we exploit the definition of the RES class provided in \cite{RES_Fang} and \cite[Ch. 3]{RichPhD}, since it is particularly useful for our aims.

\subsection{Spherically Symmetric (SS) distributions}

As a prerequisite for the definition of the RES class, Spherically Symmetric (SS) distributions need to be introduced first.
\begin{definition}
	\label{SS_vec}
	Let $\mb{z}\in \mathbb{R}^N$ be a real-valued random vector and let $\mathcal{O}$ be the set of all the orthogonal transformations such that:
	\be
	\label{orthogonal_trans}
	\begin{split}
		\mathcal{O} \ni O: \; & \mathbb{R}^N \rightarrow \mathbb{R}^N\\
		& \mb{z}\mapsto O(\mb{z})=\mb{O}\mb{z}, 
	\end{split}
	\ee         
	for any orthogonal matrix $\mb{O}$,  i.e for any $\mb{O} \in \mathbb{R}^{N \times N}$ such that $\mb{O}^T\mb{O}=\mb{O}\mb{O}^T=\mb{I}$. Then, $\mb{z}$ is said to be SS-distributed if its distribution is invariant to any orthogonal transformations $\mb{O} \in \mathcal{O}$, i.e.
	\be
	\mb{z}=_{d} \mb{O} \mb{z}.
	\ee 
	We indicate with $\mathcal{S}$ the class of all SS-distributions.
\end{definition}
As a consequence, the following properties hold true (see, e.g. \cite{RES_Fang} or \cite[Ch. 3]{RichPhD} for the proof):
\begin{itemize}
	\item[P1)] The SS-distributed random vector $\mb{z} \sim SS(g)$ has a pdf given by:
	\be
	\label{SS_pdf}
	p_{Z}(\mb{z}) = 2^{-N/2} g \left(\norm{\mb{z}}^2 \right), 
	\ee
	where $\mathcal{G} \ni g$, is a function, called \textit{density generator}, that depends on $\mb{z}$ only through $\norm{\mb{z}}$ and
	\be
	\label{set_G}
	\mathcal{G} = \lbr g: \mathbb{R}^{+}\rightarrow \mathbb{R}^{+} \left|  \int_{0}^{\infty}t^{N/2-1}g(t)dt < \infty \right. \rbr
	\ee
	where the integrability condition in \eqref{set_G} is required to guarantee the integrability of $p_{Z}(\mb{z})$ (see \cite[eq. 3.25]{RichPhD}). Consequently, the set of all SS pdfs can be described as:
	\be
	\label{S_set}
	\mathcal{S} = \lbr p_{Z}|p_{Z}(\mb{z}) = 2^{-N/2} g \left(\norm{\mb{z}}^2 \right), \forall g \in \mathcal{G} \rbr.
	\ee
	\item[P2)] Let $s_N \triangleq 2\pi^{N/2}/\Gamma(N/2)$ be the surface area of the unit sphere in $\mathbb{R}^N$, then the pdf of the random variables $\mathcal{Q}=_d\norm{\mb{z}}^2$ and $\mathcal{R}\triangleq \sqrt{\mathcal{Q}}$, called \textit{2nd-order modular variate} and \textit{modular variate} respectively \cite{Esa}, are given by:
	\be
	\label{SS_Q_pdf}
	p_{\mathcal{Q}}(q) = s_N 2^{-N/2-1} q^{N/2-1} g \left(q \right), 
	\ee
	\be
	\label{SS_R_pdf}
	p_{\mathcal{R}}(r) = s_N 2^{-N/2} r^{N-1} g \left(r^2 \right).
	\ee
	\item[P3)] \textit{The Stochastic Representation Theorem}. Let $\mb{u} \sim \mathcal{U}(\mathbb{R}S^N)$ be a random vector uniformly distributed on the real unit sphere of dimension $N$, indicated as $\mathbb{R}S^N$. If $\mb{z}\in \mathbb{R}^N$ is SS-distributed with pdf given by \eqref{SS_pdf}, then:
	\be
	\label{SRT}
	\mb{z} =_d \sqrt{\mathcal{Q}}\mb{u}=_d \mathcal{R}\mb{u},
	\ee
	where $\mathcal{Q} \sim p_{\mathcal{Q}}(q)$ in \eqref{SS_Q_pdf}, $\mathcal{R} \sim p_{\mathcal{R}}(r)$ in \eqref{SS_R_pdf}. Moreover, $\mathcal{Q}$ and $\mb{u}$ (or $\mathcal{R}$ and $\mb{u}$) are independent. Note that $\mb{u}$ in \eqref{SRT} satisfies the following three properties: $\norm{\mb{u}}=1$, $E\{\mb{u}\}=\mb{0}$ and $E\{\mb{u}\mb{u}^T\}=N^{-1}\mb{I}$.
	\item[P4)] \textit{Maximal Invariant Statistic}. The Stochastic Representation Theorem shows that there exists a one-to-one relationship between every $\mb{z} \sim SS(g)$ and every couple $(\mathcal{R}, \mb{u})$. Moreover, it is easy to verify that the modular variate $\mathcal{R}$ is a \textit{maximal invariant statistic} for the set of the SS-distributed random vectors. \footnote{For the sake of clarity, let us recall the definition of maximal invariant statistic \cite[Ch. 6]{Lehmann_TSH}. Let $\mathcal{D}=\{d\}$ be a group of one-to-one transformations on a sample space $\mathcal{X}$ and let $T$ be an invariant statistic such that $T(\mb{x}) =_d T(d(\mb{x}))$, $\forall \mb{x} \in \mathcal{X}$ and $\forall d \in \mathcal{D}$. Then, $T$ is a maximal invariant on $\mathcal{X}$ w.r.t. $\mathcal{D}$ if $T(\mb{x}_1) =_d T(\mb{x}_2)$ implies that $\mb{x}_1 =_d d (\mb{x}_2)$, $\forall \mb{x}_1,\mb{x}_2 \in \mathcal{X}$ and $\forall d \in \mathcal{D}$. Clearly, for any couple of SS-distributed random vectors $\mb{z}_1$ and $\mb{z}_2$, we have $
	\norm{\mb{z}_1} =_d \norm{\mb{z}_2} \; \Rightarrow \; \mb{z}_1 =_d \mb{O} \mb{z}_2, \quad \forall \mb{O} \in \mathcal{O}$,
	where $\mathcal{O}$ is the group of orthogonal transformations defined in \eqref{orthogonal_trans}. Consequently, $\mathcal{R} = \norm{\mb{z}}$ is a maximal invariant statistic for the set of the SS-distributed random vectors.}
\end{itemize}

We can now introduce the class of RES distributions as a \textit{semiparametric group model}.

\subsection{The RES class as a semiparametric group model}
\label{RES_section_group_mod}
At first, let us define the parameter space $\Omega \subseteq \mathbb{R}^q$ of dimension $q=N(N+3)/2$ as:
\be
\label{par_space}
\Omega = \{\bs{\phi} \in \mathbb{R}^q | \bs{\phi}=(\bs{\mu}^T,\vsigma^T)^T;\bs{\mu} \in \mathbb{R}^N,\bs{\Sigma} \in \mathcal{M}_N\},
\ee
where $\bs{\mu} \in \mathbb{R}^N$ is a real-valued $N$-dimensional vector while $\bs{\Sigma}$ is an $N \times N$ matrix belonging to the set $\mathcal{M}_N$ of all the symmetric, positive-definite matrices of dimension $N \times N$. Note that the $\mathrm{vecs}$ operator maps the $N \times N$ symmetric matrix $\bs{\Sigma}$ in an $N(N+1)/2$-dimensional vector of the entries of the lower triangular sub-matrix of $\bs{\Sigma}$ \cite{Magnus1}, \cite{Magnus2}. We can now introduce the group $\mathcal{A}$ of all affine transformations, parameterized by the parameter space $\Omega$ in \eqref{par_space}, such that
\be
\label{affine_tran}
\begin{split}
	\mathcal{A} \ni \alpha_{\bs{\phi}}: \; & \mathbb{R}^N \rightarrow \mathbb{R}^N, \; \forall \bs{\phi} \in \Omega \\
	& \mb{z}\mapsto \alpha_{\bs{\phi}}(\mb{z})= \bs{\mu} + \bs{\Sigma}^{1/2}\mb{z}.
\end{split}
\ee
The identity element $\alpha_{\bs{\phi}_e}(\cdot)$ of the group $\mathcal{A}$ is parameterized by the vector $\bs{\phi}_e=(\mb{0}^T, \mathrm{vecs}(\mb{I})^T)^T$, while the inverse transformation is given by:
\be
\label{inv_alpha}
\alpha^{-1}_{\bs{\phi}}(\cdot)= \bs{\Sigma}^{-1/2}(\cdot-\bs{\mu}).
\ee

The class of RES distributions is defined as the class of distributions that is closed under the action of the group $\mathcal{A}$ in \eqref{affine_tran} on any SS-distributed random vector. The next definition formalizes this statement.
\begin{definition}
	\label{RES_def}
	A real-valued random vector $\mb{x} \in \mathbb{R}^N$ is said to be RES-distributed with mean value $\bs{\mu}$ and scatter matrix $\bs{\Sigma}$, if it can be expressed as:
	\be
	\label{SRT_dec}
	\mb{x} = \alpha_{\bs{\phi}}(\mb{z}) = \bs{\mu} + \bs{\Sigma}^{1/2}\mb{z}=_d \bs{\mu} + \sqrt{\mathcal{Q}}\bs{\Sigma}^{1/2}\mb{u},
	\ee
	where $\bs{\phi} \in \Omega$ defined in \eqref{par_space}, $\mb{z} \sim SS(g)$ is an SS-distributed random vector, while $\mb{u} \sim \mathcal{U}(\mathbb{R}S^N)$ and the 2nd-order modular variate $\mathcal{Q}$ have been already defined in \eqref{SRT}. In particular:
	\be
	\label{Q_RES}
	\mathcal{Q}=_d\norm{\mb{z}}^2=\norm{\alpha^{-1}_{\bs{\phi}}(\mb{x})}^2 = (\mb{x}-\bs{\mu})^T\bs{\Sigma}^{-1}(\mb{x}-\bs{\mu})\triangleq Q.
	\ee 
\end{definition}
We refer the reader to \cite{RES_Fang}, \cite[Ch. 3]{RichPhD} and \cite{Esa} for the proof.

Definition \ref{RES_def} provides the link between the RES family and the semiparametric group model defined in Section \ref{semi_group_mod}. As a consequence, the explicit expression of the pdf of an RES distributed random vector can be obtained as shown in the Definition \ref{def_group_model}, Equation \eqref{semi_group_model}. In particular, the determinant of the Jacobian matrix of the inverse transformation in \eqref{inv_alpha} is $|\mb{J}(\alpha^{-1}_{\bs{\phi}})|=|\bs{\Sigma}^{-1/2}|=|\bs{\Sigma}|^{-1/2}$. Then, the pdf of any RES-distributed random vector $\mb{x}$ can be obtained from the relevant SS-distributed random vector $\mb{z}$, i.e. $p_{Z}(\mb{z})$ in \eqref{SS_pdf}, as \footnote{Note that the definition of the pdf of an RES distributed random vectors given here is consistent with the one proposed in \cite{Esa} for CES distributed random vectors. The only difference is that, in our definition, the normalizing constant $c_{N,g}$ introduced in \cite{Esa}, has been included in the density generator $g(\cdot)$.}:
	\be
	\label{RES_pdf}
	\begin{split}
 	RES_N(\mb{x};\bs{\mu},\bs{\Sigma},g)& \triangleq |\bs{\Sigma}|^{-1/2}p_{Z}(\alpha^{-1}_{\bs{\phi}}(\mb{x}))\\
		&=2^{-N/2} |\bs{\Sigma}|^{-1/2} g \left((\mb{x}-\bs{\mu})^T\bs{\Sigma}^{-1}(\mb{x}-\bs{\mu}) \right), \forall g \in \mathcal{G}. 
	\end{split}
	\ee 

Moreover, the general description of a semiparametric group model given in \eqref{semi_group_model}	can be specialized for the RES case as:
\be
\label{RES_semi_group_model}
\mathcal{P}_{\bs{\phi},g} = \lbr p_X | p_X(\mb{x}|\bs{\phi},g) = 2^{-N/2} |\bs{\Sigma}|^{-1/2} g(\norm{\alpha^{-1}_{\bs{\phi}}(\mb{x})}^2);  \bs{\phi} \in \Omega, g \in \mathcal{G} \rbr,
\ee 	
where $\mathcal{G}$ is the set of density generators given in \eqref{set_G}. Clearly, the mean vector of $\mb{x} \sim RES_N(\mb{x};\bs{\mu},\bs{\Sigma},g)$ is given by $E_0\{\mb{x}\}=\bs{\mu}$ while, if $E\{Q\}<\infty$, its covariance matrix $\mb{M}$ is $\mb{M} \triangleq E_0\{(\mb{x}-\bs{\mu})(\mb{x}-\bs{\mu})^T\} = N^{-1}E\{Q\}\mb{\Sigma}$

As extensively discussed in the literature on elliptically symmetric distributions, the representation of an RES distributed vector $\mb{x}$ is not uniquely determined by \eqref{SRT_dec}. In fact, $\mb{x} =_{d} \bs{\mu }+\sqrt{\mathcal{Q}}\bs{\Sigma}^{-1/2}\mb{u} =_{d} \bs{\mu}+\sqrt{c^{-2}\mathcal{Q}}(c\bs{\Sigma}^{-1/2})\mb{u}, \forall c>0$. This scale ambiguity can also be seen as a consequence of the functional form of an RES pdf given in \eqref{RES_pdf} since $RES_{N}(\mb{x};\bs{\mu},\bs{\Sigma},g(t))\equiv RES_{N}(\mb{x};\bs{\mu},c^2\bs{\Sigma},g(t/c^2)), \forall c>0$. To avoid this well-known identifiability problem, we impose the following constraint on the trace of $\bs{\Sigma}$, i.e.
\be
\label{cons_sigma}
c(\bs{\Sigma}) = \mathrm{tr}(\bs{\Sigma})-N = 0.
\ee
This constraint limits the parameter vector $\bs{\phi} \in \Omega$, where $\Omega$ is defined in \eqref{par_space}, in a lower dimensional smooth manifold
\be
\label{Gamma_bar}
\bar{\Omega} = \{\bs{\phi} \in \Omega | \mathrm{tr}(\bs{\Sigma})=N\},
\ee
whose dimension is $\bar{q}=q-1$. Trace constraint is only an example of all the possible constrains that can be imposed on the scatter matrix to avoid scale ambiguity. For a deep and insightful analysis of the impact of the particular constraint on $\bs{\Sigma}$ on the estimation performance, we refer to \cite{Hallin_P_2006} and \cite{PAINDAVEINE}.      

The properties of the semiparametric group model previously discussed can be exploited to derive the CSCRB for the estimation of the \textit{constrained} parameter vector $\bs{\phi}=(\bs{\mu}^T,\vsigma^T)^T \in \bar{\Omega}$, where $\bar{\Omega}$ given in \eqref{Gamma_bar}. 

\section{The Constrained Semiparametric Cram\'{e}r-Rao Bound for the RES class}
\label{SCRB_for_RES}

This section is devoted to the derivation of a closed form expression of the CSCRB for the estimation of $\bs{\phi} \in \bar{\Omega}$. The theoretical foundation of the generalization of the Cram\'{e}r-Rao inequality in the semiparametric framework can be found in \cite[Theo. 4.1]{Tsiatis}, \cite[Sec. 3.4]{BKRW}, \cite{Newey} and \cite{Begun}. Moreover, in \cite{For_EUSIPCO}, it is shown, in a tutorial manner, how the SCRB can be obtained as a result of a limit process of the classical CRB derived in the presence of a finite-dimensional nuisance parameter vector. Here, as mentioned before, we focus on the calculation of the SCRB for the particular case of the RES distributions.

As explained in the previous section, to avoid the scale ambiguity of the RES class, we need to put a constraint on the scatter matrix. In order to take this requirement into account, we propose in the sequel the extension of Theorem 1 in \cite{For_EUSIPCO} to the case of constrained, finite-dimensional, parameter vector. In particular, suppose that the finite-dimensional parameter vector of interest $\bs{\gamma}_0 \in \Gamma \subseteq \mathbb{R}^q$ is required to satisfy $k$ (with $k<q$) continuously differentiable constraints (\cite{CCRB_Stoica}, \cite{CCRB}, \cite{CMCRB}):
\be
\label{consts}
\mb{c}(\bs{\gamma}_0) = \mb{0}.
\ee
This set of constraints define a smooth manifold, of dimension $\bar{q} = q-k$, in the parameter space $\Gamma$, such that:
\be
\label{smooth_man}
\bar{\Gamma} = \lbr \bs{\gamma} \in \Gamma \subseteq \mathbb{R}^q | \mb{c}(\bs{\gamma}) = \mb{0}  \rbr
\ee 
Moreover, suppose that the $k \times q$ Jacobian matrix of the constraints, defined as $[\mb{J}_{\mb{c}}(\bs{\gamma})]_{i,j} \triangleq \partial c_i(\bs{\gamma})/ \partial \gamma_j$ has full row rank for any $\bs{\gamma} \in \Gamma$ satisfying \eqref{consts}. Consequently, there exists a matrix $\mb{U} \in \mathbb{R}^{q \times \bar{q}}$ whose columns form an orthonormal basis for the null space of $\mb{J}_{\mb{c}}(\bs{\gamma}_0)$, i.e.
\be
\label{U_prop}
\mb{J}_{\mb{c}}(\bs{\gamma}_0)\mb{U}=\mb{0}_{k \times \bar{q}}, \qquad \mb{U}^T\mb{U}=\mb{I}.
\ee
\begin{theorem}
	\label{Theo_CSCRB}	
	The Constrained Semiparametric Cram\'{e}r-Rao Bound (CSCRB) for the estimation of the constrained finite-dimensional vector $\bs{\gamma}_0 \in \bar{\Gamma}$ in the presence of the nuisance function $l_0\in \mathcal{L}$ is given by:
	\be
	\label{SEB}
	\mathrm{CSCRB}(\bs{\gamma}_0|l_0) = \mb{U}(\mb{U}^T\bar{\mb{I}}(\bs{\gamma}_0|l_0)\mb{U})^{-1}\mb{U}^T,
	\ee
	where:
	\be
	\label{S_E_FIM}
	\bar{\mb{I}}(\bs{\gamma}_0|l_0) \triangleq E_0\{\bar{\mb{s}}_0(\bar{\mb{s}}_0)^T\},
	\ee
	is the \textit{semiparametric Fisher Information Matrix} (SFIM) and $\bar{\mb{s}}_0$ is the \textit{semiparametric efficient score vector} defined as:
	\be
	\label{Semi_eff_score}
	\bar{\mb{s}}_0 = \mb{s}_{\bs{\gamma}_0} - \Pi(\mb{s}_{\bs{\gamma}_0}|\mathcal{T}_{l_0}),
	\ee
	where $\Pi(\mb{s}_{\bs{\gamma}_0}|\mathcal{T}_{l_0})$ is the orthogonal projection of the score vector of the parameters of interest $\mb{s}_{\bs{\gamma}_0}$ on the semiparametric nuisance tangent space. Finally, matrix $\mb{U}$ is defined in \eqref{U_prop}. 
	Note that $\bar{\mb{s}}_0$, $\mb{s}_{\bs{\gamma}_0}$ and $\Pi(\mb{s}_{\bs{\gamma}_0}|\mathcal{T}_{l_0})$ are $q$-dimensional functions of the observation vector $\mb{x}$.
\end{theorem}

The proof of the \virg{unconstrained} part of this theorem can be found in \cite[Theo. 4.1]{Tsiatis} and in \cite{Newey}, while a more abstract and general formulation can be found in \cite{Begun} and in \cite[Sec. 3.4]{BKRW}. The proof of the \virg{constrained} part can be obtained through a straightforward application of the approach discussed in \cite{CCRB} for the constrained CRB. 

The remainder of this section is devoted to the evaluation of the CSCRB in \eqref{SEB} for the estimation of the mean vector $\bs{\mu}$ and of the constrained scatter matrix $\bs{\Sigma}$, such that $\mathrm{tr}(\bs{\Sigma})=N$, of a RES distributed random vector $\mb{x} \sim RES_N(\mb{x};\bs{\mu}_0,\bs{\Sigma}_0,g_0)$ in the presence of the nuisance function $g_0 \in \mathcal{G}$. To this end, according to Theorem \ref{Theo_CSCRB}, we have to evaluate:
\begin{itemize}
	\item [\textit{A}.] the score vector of the parameters of interest $\mb{s}_{\bs{\phi}_0}\equiv\mb{s}_{\bs{\phi}_0}(\mb{x})$, where $\bs{\phi}_0=(\bs{\mu}_0^T,\vsigmat^T)^T \in \Omega$ given in \eqref{par_space},
	\item [\textit{B}.] the projection operator $\Pi(\mb{s}_{\bs{\phi}_0}|\mathcal{T}_{g_0})$,
	where $\mathcal{T}_{g_0}\equiv \mathcal{T}_{g_0}(\bs{\phi}_0)$ is the semiparametric nuisance tangent space for the RES distributions class evaluated at the true semiparametric vector $(\bs{\phi}_0^T,g_0)^T$,
	\item [\textit{C}.]	the semiparametric efficient score vector $\bar{\mb{s}}_0\equiv \bar{\mb{s}}_0(\mb{x})$ in \eqref{Semi_eff_score} and the SFIM $\bar{\mb{I}}(\bs{\phi}_0|g_0)$ in \eqref{S_E_FIM},
	\item [\textit{D}.] the matrix $\mb{U}$ in \eqref{U_prop} and then the CSCRB in \eqref{SEB}.
\end{itemize}

In what follows, we will provide this calculation step-by-step. 

\subsection{Evaluation of the score vector $\mb{s}_{\bs{\phi}_0}(\mb{x})$}
\label{eval_score_vec}
By definition, the score vector of the parameters of interest is given by:
\be
\mb{s}_{\bs{\phi}_0}(\mb{x})=\nabla_{\bs{\phi}}\ln p_X(\mb{x}|\bs{\phi}_0,g_0) = \left( \begin{array}{c}
	\mb{s}_{\bs{\mu}_0}(\mb{x})\\
	\mb{s}_{\mathrm{vecs}(\bs{\Sigma}_0)}(\mb{x})\end{array}\right), 
\ee 
where $\bs{\phi}_0 \in \Omega$ and $\Omega$ is the \textit{unconstrained} parameter space defined in \eqref{par_space} and:
\be
\label{grad_mu}
\mb{s}_{\bs{\mu}_0}(\mb{x}) = \nabla_{\bs{\mu}}\ln p_X(\mb{x}|\bs{\mu}_0,\bs{\Sigma}_0,g_0),
\ee
\be
\label{grad_sigma}
\mb{s}_{\mathrm{vecs}(\bs{\Sigma}_0)}(\mb{x}) = \nabla_{\mathrm{vecs}(\bs{\Sigma})}\ln p_X(\mb{x}|\bs{\mu}_0,\bs{\Sigma}_0,g_0),
\ee
and where $\bs{\mu}_0$, $\bs{\Sigma}_0$ and $g_0$ represents the true mean vector, the true scatter matrix and the true density generator, respectively. By substituting in \eqref{grad_mu} the explicit expression of $p_X(\mb{x}|\bs{\mu}_0,\bs{\Sigma}_0,g_0)$ given in \eqref{RES_pdf} and by exploiting the differentiation rules with respect to vector and matrices provided e.g. in \cite[Ch. 8]{Magnus}, we have that:
\be
\label{score_mu}
\begin{split}
	\mb{s}_{\bs{\mu}_0}(\mb{x}) &= -2 \psi_0(Q_0) \bs{\Sigma}_0^{-1} (\mb{x}-\bs{\mu}_0)\\
	&=_d -2 \sqrt{\mathcal{Q}} \psi_0(\mathcal{Q}) \bs{\Sigma}_0^{-1/2} \mb{u}\\
\end{split}
\ee
where, according to \eqref{Q_RES}, 
\be
Q_0 = \norm{\alpha^{-1}_{\bs{\phi}_0}(\mb{x})}^2 = (\mb{x}-\bs{\mu}_0)^T\bs{\Sigma}_0^{-1}(\mb{x}-\bs{\mu}_0)=_d \mathcal{Q},
\ee
and the last two equalities follow directly from the Stochastic Representation of a RES vector given in \eqref{SRT_dec} and 
\be
\label{psi}
\psi_0(t) \triangleq \frac{1}{g_0(t)}\frac{dg_0(t)}{dt}.
\ee
Similarly, the term $\mb{s}_{\vsigmat}(\mb{x})$ in \eqref{grad_sigma} can be evaluated by applying the rules of the differential matrix calculus detailed in \cite[Ch. 8]{Magnus} and by using standard properties of the Kronecker product and of the vec operator (\cite[Ch. 15]{Magnus}, \cite{Magnus1,Magnus2}) as:    
\be
\label{score_sigma}
\begin{split}
	\mb{s}_{\vsigmat}(\mb{x}) &= -\mb{D}_N^T\left( 2^{-1}\vcsigmatinv + \psi_0(Q_0)  \kronsigmatinv \mathrm{vec}((\mb{x}-\bs{\mu}_0)(\mb{x}-\bs{\mu}_0)^T)\right) \\
	& =_d -\mb{D}_N^T\left( 2^{-1}\vcsigmatinv + \mathcal{Q}\psi_0(\mathcal{Q})  \kronsigmatinvm \mathrm{vec}(\mb{u}\mb{u}^T)\right)\\
\end{split}
\ee
As before, the second equality follows from the Stochastic Representation \eqref{SRT_dec} while $\mb{D}_N$ is the $N^2 \times N(N+1)/2$ \textit{duplication matrix} that is implicitly defined by the equality $\mb{D}_N\mathrm{vecs}(\mb{A})=\mathrm{vec}({\mb{A}})$ for any $N^2 \times N^2$ symmetric matrix $\mb{A}$ (\cite{Magnus1}, \cite{Magnus2}). See also \cite{Grec1, Grec3, Grec2} for similar calculation. 

\subsection{Evaluation of the projection operator $\Pi(\mb{s}_{\bs{\phi}_0}|\mathcal{T}_{g_0})$}
\label{eval_proj_op}
In order to obtain the explicit expression of the projection operator, we will exploit the fact that the RES class is a semiparametric group model. In particular, $\Pi(\mb{s}_{\bs{\phi}_0}|\mathcal{T}_{g_0})$ can be obtained by specializing Proposition \ref{semi_group_tan_proj} for the RES distributions. 

For the sake of clarity, let us recall that, according to the definition of the group $\mathcal{A}$ of the affine tranformations \eqref{affine_tran}, for any $\bs{\phi}_0 \in \Omega$, we have:
\be
\label{alfa_d}
\alpha_{\bs{\phi}_0}(\mb{z}) \triangleq \bs{\mu}_0 + \bs{\Sigma}_0^{1/2} \mb{z}, \quad \mb{z} \sim SS(g_0),
\ee
\be
\label{alfa_identity}
\alpha_{\bs{\phi}_e}(\mb{z}) \triangleq \mb{0} + \mb{I}^{1/2} \mb{z} = \mb{z}, \quad \mb{z} \sim SS(g_0),
\ee
\be
\label{alfa_inv}
\alpha_{\bs{\phi}_0}^{-1}(\mb{x}) \triangleq \bs{\Sigma}_0^{1/2} (\mb{x}-\bs{\mu}_0), \quad \mb{x} \sim RES_{N}(\mb{x};\bs{\mu}_0,\bs{\Sigma}_0,g_0).
\ee

Let $\mathcal{T}_{g_0} \equiv \mathcal{T}_{g_0}(\bs{\phi}_0)$ and $\mathcal{T}_{\mathcal{S}_0} \equiv \mathcal{T}_{g_0}(\bs{\phi}_e)$ be the semiparametric nuisance tangent spaces of the RES class evaluated at the true semiparametric vector $(\bs{\phi}_0^T,g_0)^T$ and at $(\bs{\phi}_e^T,g_0)^T$, where $\bs{\phi}_e=(\mb{0}^T,\mathrm{vecs}(\mb{I})^T)^T$ is the vector that characterizes the identity transformation given in \eqref{alfa_identity}. Note that, since it is evaluated at the identity transformation, $\mathcal{T}_{\mathcal{S}_0}$ can be interpreted as the tangent space of the SS distribution class evaluated at the true $g_0$, and this explains the chosen notation. Now, by directly applying Proposition \ref{semi_group_tan_proj}, we have that the projection of the score vector of the parameters of interest $\mb{s}_{\bs{\phi}_0}(\mb{x})$ on the tangent space $\mathcal{T}_{g_0}$ can be expressed as: 
\be
\label{proj_calc}
(\Pi(\mb{s}_{\bs{\phi}_0}(\mb{x})|\mathcal{T}_{g_0}))(\mb{x}) = (\Pi((\mb{s}_{\bs{\phi}_0} \circ \alpha_{\bs{\phi}_0})(\mb{z})|\mathcal{T}_{\mathcal{S}_0})\circ \alpha^{-1}_{\bs{\phi}_0})(\mb{x}),
\ee
where $\mb{x} \sim RES_{N}(\mb{x};\bs{\mu}_0,\bs{\Sigma}_0,g_0)$ and $\mb{z} \sim SS(g_0)$.

In what follows, we show how to obtain an explicit expression of the projection in \eqref{proj_calc}.

\subsubsection{Calculation of $(\mb{s}_{\bs{\phi}_0}\circ \alpha_{\bs{\phi}_0})(\mb{z})$} 

From a visual inspection of the expressions in \eqref{score_mu} and \eqref{score_sigma}, we notice that they can be rewritten as a function of the inverse transformation $\alpha_{\bs{\phi}_0}^{-1}(\mb{x})$, i.e.:
\be
\label{score_mu_alfa}
\mb{s}_{\bs{\mu}_0}(\mb{x}) =(\tilde{\mb{s}}_{\bs{\mu}_0} \circ \alpha_{\bs{\phi}_0}^{-1})(\mb{x})= -2 \psi_0(Q_0) \bs{\Sigma}_0^{-1/2} \alpha_{\bs{\phi}_0}^{-1}(\mb{x}),
\ee
and
\be
\label{score_sigma_alfa}
\begin{split}
	\mb{s}_{\vsigmat}(\mb{x}) &= (\tilde{\mb{s}}_{\vsigmat}\circ \alpha_{\bs{\phi}_0}^{-1})(\mb{x}) \\
	&= -\mb{D}_N^T\left( 2^{-1}\vcsigmatinv + \psi_0(Q_0)  \kronsigmatinvm \mathrm{vec}(\alpha_{\bs{\phi}_0}^{-1}(\mb{x})\alpha_{\bs{\phi}_0}^{-1}(\mb{x})^T)\right).
\end{split}
\ee
Then, we have that:
\be
(\mb{s}_{\bs{\phi}_0} \circ  \alpha_{\bs{\phi}_0})(\mb{z}) = \left( \begin{array}{c}
	(\tilde{\mb{s}}_{\bs{\mu}_0}\circ \alpha_{\bs{\phi}_0}^{-1} \circ \alpha_{\bs{\phi}_0})(\mb{z})\\
	(\tilde{\mb{s}}_{\mathrm{vecs}(\bs{\Sigma}_0)}\circ \alpha_{\bs{\phi}_0}^{-1} \circ \alpha_{\bs{\phi}_0})(\mb{z})\end{array}\right) = \left( \begin{array}{c}
	\tilde{\mb{s}}_{\bs{\mu}_0} (\mb{z})\\
	\tilde{\mb{s}}_{\mathrm{vecs}(\bs{\Sigma}_0)}(\mb{z})\end{array}\right),
\ee 
and
\be
\label{score_mu_alfa_z}
\tilde{\mb{s}}_{\bs{\mu}_0}(\mb{z}) = -2 \psi_0(Q_0) \bs{\Sigma}_0^{-1/2} \mb{z} =_d -2 \sqrt{\mathcal{Q}}\psi_0(\mathcal{Q}) \bs{\Sigma}_0^{-1/2} \mb{u}
\ee
\be
\label{score_sigma_alfa_z}
\begin{split}
	\tilde{\mb{s}}_{\vsigmat}(\mb{z})&= -\mb{D}_N^T\left( 2^{-1}\vcsigmatinv + \psi_0(Q_0)  \kronsigmatinvm \mathrm{vec}(\mb{z}\mb{z}^T)\right)\\
	&=_d-\mb{D}_N^T\left( 2^{-1}\vcsigmatinv + \mathcal{Q}\psi_0(\mathcal{Q})  \kronsigmatinvm \mathrm{vec}(\mb{u}\mb{u}^T)\right).
\end{split}
\ee

\subsubsection{Derivation of $\mathcal{T}_{\mathcal{S}_0}$} 
The next step is the evaluation of $\mathcal{T}_{\mathcal{S}_0}$, i.e. the tangent space of $\mathcal{S}$ in \eqref{S_set}, evaluated at the true density generator $g_0$. Using the procedure discussed in the Appendix A.3 of \cite{BKRW}, it is possible to verify that $\mathcal{T}_{\mathcal{S}_0}$ is a $q$-replicating Hilber space $\mathcal{T}_{\mathcal{S}_0} = \mathcal{T} \times \ldots \times \mathcal{T}$ (see also Appendix \ref{Hilber_random}) such that: 
\be
\mathcal{T}_{\mathcal{S}_0} \triangleq \{l\mb{a}|\; \mb{a} \mathrm{ \; is \; any \;vector \;in  \;} \mathbb{R}^{q}, l \in \mathcal{T}\},
\ee
where
\be
\begin{split}
	\mathcal{T} &= \lbr l \in \mathcal{H}^1 | l \mathrm{ \; is \; invariant \;under \;}\mathcal{O} \rbr \\
	&=\lbr l | l(\mb{z}) \equiv l(\norm{\mb{z}}), \;  E_0\{l(\norm{\mb{z}}\} =0\rbr,	
\end{split}
\ee
and $\mathcal{O}$ is the group of orthogonal transformations defined in \eqref{orthogonal_trans}. Let us now recall that, according to Property P4 in Sec. \ref{RES_section}, the modular variate $\mathcal{R}$ is a maximal invariant statistic for an SS distribution. Then, by using the procedure discussed in \cite[Sec. 6.3, Example 1]{BKRW} and in accordance with the discussion provided in Appendix \ref{cond_expt_sect}, we have that the projection operator $\Pi(\cdot|\mathcal{T}_{\mathcal{S}_0})$ on the tangent space $\mathcal{T}_{\mathcal{S}_0}$ can be obtained as the expectation operator $E_{0|\mathcal{R}}\{\cdot|\mathcal{R}\}$ with respect to the maximal invariant statistic $\mathcal{R}$, i.e.
\be
\label{proj_form_l}
\Pi(\mb{l}|\mathcal{T}_{\mathcal{S}_0}) = E_{0|\mathcal{R}}\{\mb{l}|\mathcal{R}\}, \quad \forall \mb{l} \in \mathcal{H}^q.
\ee

\subsubsection{Calculation of the projection $\Pi(\mb{s}_{\bs{\phi}_0}|\mathcal{T}_{g_0})$}
In order to evaluate $\Pi(\cdot|\mathcal{T}_{g_0})$, we can use the property of the semiparametric group models given in \eqref{proj_calc}. Let us start by deriving the expression for $\Pi((\mb{s}_{\bs{\phi}_0} \circ \alpha_{\bs{\phi}_0})|\mathcal{T}_{\mathcal{S}_0}) \equiv \Pi(\tilde{\mb{s}}_{\bs{\phi}_0}|\mathcal{T}_{\mathcal{S}_0})$. By exploiting the results collected in the previous subsections, $\Pi(\tilde{\mb{s}}_{\bs{\phi}_0}|\mathcal{T}_{\mathcal{S}_0})$ can be easily obtained by substituting \eqref{score_mu_alfa_z} and \eqref{score_sigma_alfa_z} in \eqref{proj_form_l} and then evaluating the expectation operator. Specifically:
\be
\label{proj_mean_val}
\Pi(\tilde{\mb{s}}_{\bs{\mu}_0}|\mathcal{T}_{\mathcal{S}_0}) = E_{0|\mathcal{R}}\{\tilde{\mb{s}}_{\bs{\mu}_0}|\mathcal{R}\} =_d -2 \sqrt{\mathcal{Q}}\psi_0(\mathcal{Q}) \bs{\Sigma}_0^{-1/2} E\{\mb{u}\} = \mb{0},
\ee
and
\be
\begin{split}
	\Pi(\tilde{\mb{s}}_{\vsigmat}&|\mathcal{T}_{\mathcal{S}_0}) = E_{0|\mathcal{R}}\{\tilde{\mb{s}}_{\vsigmat}(\mb{z})|\mathcal{R}\} \\
	& =_d -\mb{D}_N^T\left( 2^{-1}\vcsigmatinv + \mathcal{Q}\psi_0(\mathcal{Q})  \kronsigmatinvm E\{\mathrm{vec}(\mb{u}\mb{u}^T)\}\right)\\
	& = -\mb{D}_N^T\left( \frac{1}{2} + \frac{1}{N} \mathcal{Q}\psi_0(\mathcal{Q}) \right)\vcsigmatinv,
\end{split}
\ee
then, consequently:
\be
\label{summariz}
\Pi(\tilde{\mb{s}}_{\bs{\phi}_0}|\mathcal{T}_{\mathcal{S}_0}) =_d \left( \begin{array}{c}
	\mb{0}\\
	-\mb{D}_N^T\left( \frac{1}{2} + \frac{1}{N} \mathcal{Q}\psi_0(\mathcal{Q}) \right)\vcsigmatinv\end{array}\right)=_d \Pi(\mb{s}_{\bs{\phi}_0}|\mathcal{T}_{g_0}),
\ee 
where the last equality follows from \eqref{proj_calc} and from the fact that $\Pi(\tilde{\mb{s}}_{\bs{\phi}_0}|\mathcal{T}_{\mathcal{S}_0})$ does not depend on $\mb{z}$.

Now, a comment is in order. Equation \eqref{summariz} tells us that the score function of the mean value is orthogonal to the nuisance tangent space $\mathcal{T}_{g_0}$. This means that not knowing the true density generator does not have any impact in the (asymptotic) estimation performance of the mean vector $\bs{\mu}_0$ \cite{Bickel_paper}.

\subsection{Calculation of the semiparametric efficient score vector $\bar{\mb{s}}_0(\mb{x})$ and of the  SFIM $\bar{\mb{I}}(\bs{\phi}_0,g_0)$}
By collecting the previous results, the semiparametric efficient score vector in \eqref{Semi_eff_score} can be evaluated as:
\be
\label{eff_score_res}
\begin{split}
	\bar{\mb{s}}_0(\mb{x}) &= \mb{s}_{\bs{\phi}_0}(\mb{x}) - \Pi(\mb{s}_{\bs{\phi}_0}(\mb{x})|\mathcal{T}_{g_0})(\mb{x}) = \left( \begin{array}{c}
		\bar{\mb{s}}_{\bs{\mu}_0}(\mb{x})\\
		\bar{\mb{s}}_{\vsigmat}(\mb{x})\end{array}\right)\\ 
	&=_d \left( \begin{array}{c}
		-2 \sqrt{\mathcal{Q}} \psi_0(\mathcal{Q}) \bs{\Sigma}_0^{-1/2} \mb{u}\\
		-\mb{D}_N^T\mathcal{Q}\psi_0(\mathcal{Q})\left( \kronsigmatinvm \mathrm{vec}(\mb{u}\mb{u}^T) - \frac{1}{N}\vcsigmatinv \right)\end{array}\right).
\end{split}
\ee
Finally, through direct calculation, the SFIM $\bar{\mb{I}}(\bs{\phi}_0,g_0)$ can be obtained as:
\be
\label{SFIM}
\bar{\mb{I}}(\bs{\phi}_0|g_0)=E_0\{\bar{\mb{s}}_0(\mb{x})\bar{\mb{s}}_0(\mb{x})^T\} = \left( \begin{array}{cc}
	\mb{C}_0(\bar{\mb{s}}_{\bs{\mu}_0}) & \mb{0} \\
	\mb{0}^T   &  \mb{C}_0(\bar{\mb{s}}_{\vsigmat})
\end{array}\right),
\ee
where $\mb{C}_0(\mb{l})\triangleq E_0\{\mb{l}\mb{l}^T\} \forall \mb{l} \in \mathcal{H}^q$. Note that the off-diagonal block matrices in \eqref{SFIM} are nil because all the third-order moments of $\mb{u}$ vanish \cite[Lemma 1]{Esa}. From the previous results and using some algebra, we get:
\be
\mb{C}_0(\bar{\mb{s}}_{\bs{\mu}_0}) = \frac{4E\{\mathcal{Q}\psi_0(\mathcal{Q})^2\}}{N} \bs{\Sigma}_0^{-1}, 
\ee
and
\be
\label{cov_mat_eff_scat}
\mb{C}_0(\bar{\mb{s}}_{\vsigmat}) = \frac{2E\{\mathcal{Q}^2\psi_0(\mathcal{Q})^2\}}{N(N+2)} \mb{D}_N^T \left( \kronsigmatinv - \frac{1}{N}\vcsigmatinv \vcsigmatinv^T \right)\mb{D}_N.
\ee
Note that we had not imposed the constraint on the scatter matrix $\bs{\Sigma}_0$ as yet. In particular, in all the previous equations, $\bs{\Sigma}_0$ can be considered as the \textit{unconstrained} scatter matrix. The next subsection is then dedicated to the derivation of the SCRB for the constrained parameter vector.

\subsection{Evaluation of the $\mathrm{CSCRB}(\bs{\phi}_0|g_0)$}
\label{const_SCRB}
As showed in Theorem \ref{Theo_CSCRB}, as a prerequisite of the derivation of the CSCRB on the estimation of $\bs{\phi}_0 \in \bar{\Omega}$, we have to calculate the matrix $\mb{U}$ defined in \eqref{U_prop}. This can be done by using the same procedure discussed in \cite{Fortunati2016}. Specifically, let us start by evaluating the gradient of the constraint in \eqref{cons_sigma} as:
\be
\mb{J}_{c}(\bs{\Sigma}_0)=\nabla^T_{\vsigma}c(\bs{\Sigma}_0) = \mb{1}^T_{I},
\ee
where $\mb{1}_{I}$ is the $N(N+1)/2$-dimensional column vector defined as:
\be
[\mb{1}_{I}]_i = \left\lbrace \begin{array}{cc}
	1 & i \in I\\
	0 & \mathrm{otherwise}
\end{array}\right. 
\ee
where $I = \lbr i\left| i=1+N(j-1)-(j-1)(j-2)/2,j=1,\ldots,N \right. \rbr$.

Then, $\mb{U}$ can be obtained by numerically evaluating, using singular value decomposition (SVD), the $\bar{q}=q-1$ orthonormal eigenvectors associated with the zero eigenvalue of $\mb{1}_{I}$.

Finally, the \textit{constrained} SCRB (CSCRB) for the estimation of $\bs{\phi}_0 \in \bar{\Omega}$ in \eqref{Gamma_bar} can be expressed as:
\be
\label{SCRB_gun}
\mathrm{CSCRB}(\bs{\phi}_0|g_0) = \left( \begin{array}{cc}
	\frac{N}{4E\{\mathcal{Q}\psi_0(\mathcal{Q})^2\}} \bs{\Sigma}_0 & \mb{0} \\
	\mb{0}^T   &  \mb{U}\left( \mb{U}^T\mb{C}_0(\bar{\mb{s}}_{\vsigmat})\mb{U}\right)^{-1}\mb{U}^T
\end{array}\right).
\ee

Note that the block-diagonal structure of $\mathrm{CSCRB}(\bs{\phi}_0|g_0)$ implies that not knowing the mean vector $\bs{\mu}_0$ does not have any impact on the asymptotic performance in the estimation of the scatter matrix $\bs{\Sigma}_0$. From a practical point of view, this also means that the unknown $\bs{\mu}_0$ can be substituted with any consistent estimator without affecting the asymptotically optimal performance of the scatter matrix estimator.

\section{Simulation results for RES distributed data}
\label{simulation_RES}
In this section, we investigate the efficiency of three well-known scatter matrix estimators with respect to the CSCRB: the constrained Sample Covariance Matrix (CSCM) estimator, the constrained Tyler's (C-Tyler) estimator and the constrained Huber's (C-Hub) estimator. Note that none of these estimators relies on the a-priori knowledge of the true density generator $g_0 \in \mathcal{G}$.

Assume to have a set of $M$, RES-distributed, observation vectors $\{\mb{x}_m \}_{m=1}^M$. Let us define $\{\bar{\mb{x}}_m \}_{m=1}^M$ as the set of $M$ vectors such that:
\be
\bar{\mb{x}}_m = \mb{x}_m - \hat{\bs{\mu}}, \quad m=1,\ldots,M
\ee
and $\hat{\bs{\mu}}$ is the sample mean estimator, i.e. $\hat{\bs{\mu}} \triangleq N^{-1} \sum\nolimits_{m=1}^{M} \mb{x}_m$.

The CSCM estimate can then be expressed as \cite{Pas}: 
\be
\label{CSCM}
\hat{\bs{\Sigma}}_{CSCM} \triangleq \frac{N}{\mathrm{tr}(\bs{\Sigma}_{SCM})} \bs{\Sigma}_{SCM}, \quad \bs{\Sigma}_{SCM} \triangleq \frac{1}{M} \sum_{m=1}^{M} \bar{\mb{x}}_m\bar{\mb{x}}_m^T,
\ee 
while C-Tyler and C-Hub estimates are the convergence points of the following iterative algorithm: 
\be
\label{C_Tyler}
\left\lbrace 
\begin{aligned}
	\mb{S}_T^{(k+1)} & = \frac{1}{M}\sum_{m=1}^{M} \varphi(t^{(k)})\bar{\mb{x}}_m\bar{\mb{x}}_m^T\\
	\hat{\bs{\Sigma}}_T^{(k+1)} &= N \mb{S}_T^{(k+1)}/\mathrm{tr}(	\mb{S}_T^{(k+1)})
\end{aligned} 
\right. ,
\ee
where $t^{(k)}=\bar{\mb{x}}_m^T(\hat{\bs{\Sigma}}_{T}^{(k)})^{-1}\bar{\mb{x}}_m$ and the starting point is $\hat{\bs{\Sigma}}_{T}^{(0)} = \mb{I}$. The weight function $\varphi(t)$ for Tyler's estimator is defined as (see e.g. \cite{Esa,Chit} and \cite[Ch. 4]{book_zoubir} and references therein):
\be
\label{w_Tyler}
\varphi_{Tyler}(t)=N/t,
\ee 
whereas the weight function for Huber's estimator is given by (\cite{Esa}, \cite{Ollila_Huber} and \cite[Ch. 4]{book_zoubir}):
\be
\label{w_Huber}
\varphi_{Hub}(t)= \lbr
\begin{array}{cc}
	1/b & t\leqslant \delta^2 \\
	\delta^2/(tb) & t> \delta^2
\end{array} \right.,
\ee 
and $\delta = F_{\chi_N^2}(u)$ where $F_{\chi_N^2}(\cdot)$ indicates the distribution of a chi-squared random variable with $N$ degrees of freedom and $q \in (0,1]$ is a tuning parameter. Moreover, the parameter $b$ is usually chosen as $b=F_{\chi_{N+2}^2}(\delta^2)+\delta^2(1-F_{\chi_N^2}(\delta^2))/N$ \cite{Esa}, \cite{Ollila_Huber}.

To compare the three scatter matrix estimators with the CSCRB, we define the following performance index:
\be
\varepsilon_\alpha \triangleq \norm{E\{(\mathrm{vecs}(	\hat{\bs{\Sigma}}_\alpha)-\mathrm{vecs}(\bs{\Sigma}_0))(\mathrm{vecs}(	\hat{\bs{\Sigma}}_\alpha)-\mathrm{vecs}(\bs{\Sigma}_0))^T\}}_F,
\ee
where $\alpha = \{CSCM,C-Tyler,C-Hub(u)\}$ indicates the particular estimator under test and $\norm{\mb{A}}^2_F \triangleq \mathrm{tr}(\mb{A}^T\mb{A})$ is the Frobenius norm of matrix $\mb{A}$. Similarly, for the sample mean, we have the following index:
\be
\varepsilon_{\bs{\mu}_0} \triangleq \norm{E\{(\hat{\bs{\mu}}-\bs{\mu}_0)(	\hat{\bs{\mu}}-\bs{\mu}_0)^T\}}_F.
\ee
For the sake of comparison, we show in the following figures also the \textit{constrained} CRB (CCRB) for the estimation of $\bs{\phi}_0=(\bs{\mu}_0^T,\vsigmat^T)^T \in \bar{\Omega}$. The classical FIM can be obtained from the score vectors previously derived in Subsection \ref{eval_score_vec} as a block matrix of the form \cite{Grec3}: 
\be
\label{FIM_gk}
\mb{I}(\bs{\phi}_0) = E_0\{\mb{s}_{\bs{\phi}_0}\mb{s}^T_{\bs{\phi}_0}\} = \left( \begin{array}{cc}
	\mb{C}_0(\mb{s}_{\bs{\mu}_0}) & \mb{0} \\
	\mb{0}^T   &  \mb{C}_0(\mb{s}_{\vsigmat})
\end{array}\right).
\ee
Through direct calculation, it is easy to verify that:
\be
\label{C_mu_p}
\mb{C}_0(\mb{s}_{\bs{\mu}_0}) = \mb{C}_0(\bar{\mb{s}}_{\bs{\mu}_0}) =  \frac{4E\{\mathcal{Q}\psi_0(\mathcal{Q})^2\}}{N} \bs{\Sigma}_0^{-1},
\ee  
and
\be
\label{FIM_Sigma_gk}
\mb{C}_0(\mb{s}_{\vsigmat}) = \mb{D}_N^T\left( a_1 \vcsigmatinv\vcsigmatinv^T +  a_2 \kronsigmatinv  \right)\mb{D}_N
\ee
where:
\be
\label{a1}
a_1 \triangleq \frac{1}{4} + \frac{E\{\mathcal{Q}\psi_0(\mathcal{Q})\}}{N} + \frac{a_2}{2},
\ee
\be
\label{a2}
a_2 \triangleq \frac{2E\{\mathcal{Q}^2\psi_0(\mathcal{Q})^2\}}{N(N+2)}.
\ee
It is worth highlighting here that the difference between the classical FIM in \eqref{FIM_gk} and the SFIM in \eqref{SFIM} is due to the different expressions of the covariance matrix of the efficient score vector $\bar{\mb{s}}_{\vsigmat}$ in \eqref{cov_mat_eff_scat} and of the one of the score vector $\mb{s}_{\vsigmat}$ in \eqref{FIM_Sigma_gk}.

Finally, the CCRB on the estimation of $\bs{\phi}_0 \in \bar{\Omega}$ can be obtained using exactly the same procedure discussed for the CSCRB in Subsection \ref{const_SCRB} (see also \cite{Fortunati2016}).

As performance bounds, the following indices are plotted:
\be
\varepsilon_{CCRB,\bs{\Sigma}_0} \triangleq \norm{[\mathrm{CCRB}(\bs{\phi}_0)]_{\bs{\Sigma}_0}}_F,
\ee
\be
\varepsilon_{CSCRB,\bs{\mu}_0} \triangleq \norm{[\mathrm{CSCRB}(\bs{\phi}_0,g_0)]_{\bs{\mu}_0}}_F,
\ee
\be
\varepsilon_{CSCRB,\bs{\Sigma}_0} \triangleq \norm{[\mathrm{CSCRB}(\bs{\phi}_0,g_0)]_{\bs{\Sigma}_0}}_F,
\ee
where $[\cdot]_{\bs{\mu}_0}$ and $[\cdot]_{\bs{\Sigma}_0}$ indicate the top-left and the bottom-right submatrices of the $\mathrm{CCRB}(\bs{\phi}_0)$ and of the $\mathrm{CSCRB}(\bs{\phi}_0,g_0)$, respectively.

We analyze two different cases:
\begin{enumerate}
	\item The true RES distribution is a \textit{t}-distribution,
	\item The true RES distribution is a Generalized Gaussian (GG) distribution.
\end{enumerate}

The simulation parameters that are common to the two cases are:
\begin{itemize}
	\item $[\bs{\Sigma}_0]_{i,j} = \rho^{|i-j|}, \; i,j=1,\ldots,N$. Moreover, $\rho = 0.8$ and $N=8$.
	\item The data power is chosen to be $\sigma_X^2 = E_\mathcal{Q}\{\mathcal{Q}\}/N = 4$.
	\item The data mean value is chosen to be $[\bs{\mu}_0]_i=1, \; i=1,\ldots,N$.
	\item The number of the available independent and identically distributed (i.i.d.) data vectors is $M=3N=24$. Note that, since we assume to have $M$ i.i.d. data vectors, the SFIM in \eqref{SFIM} and the FIM in \eqref{FIM_gk} have to be multiply by $M$.
	\item The tuning parameter $u$ of Huber's estimator has been chosen as $u=0.9,0.5,0.1$. Note that for $u=1$ Huber's estimator is equal to the SCM, while for $u\rightarrow 0$ Huber's estimator tends to Tyler's estimator \cite{Esa}.
	\item The number of independent Monte Carlo runs is $10^6$.
\end{itemize}

\subsection{Case 1: the $t$-distribution}
The density generator for the $t$-distribution is:
\be
\label{g_t}
g_0(t) \triangleq \frac{2^{N/2}\Gamma(\frac{\lambda+N}{2})}{\pi^{N/2}\Gamma({\lambda}/2 )}\left( \frac{\lambda}{\eta}\right) ^{{\lambda}/2}\left( \frac{\lambda}{\eta} + t \right)^{-\frac{\lambda+N}{2}} 
\ee
and then $\psi_0(t)$ in \eqref{psi} is given by $\psi_0(t)=-2^{-1}(\lambda + N)(\lambda/\eta + t)^ {-1}$. Consequently, from \eqref{SS_Q_pdf}, we have that:
\be
p_\mathcal{Q}(q) =  \frac{\Gamma(\frac{\lambda+N}{2})}{\Gamma({\lambda}/2 )\Gamma(N/2)}\left( \frac{\lambda}{\eta}\right) ^{{\lambda}/2} q^{N/2-1} \left( \frac{\lambda}{\eta} + q \right)^{-\frac{\lambda+N}{2}}
\ee

Using the integral in \cite[pp. 315, n. 3.194 (3)]{Integrals}, we have that:
\be
E\{\mathcal{Q}\psi_0(\mathcal{Q})\} = -N/2
\ee
\be
E\{\mathcal{Q}\psi_0(\mathcal{Q})^2\}  =  \frac{\eta N (\lambda + N)}{4(N +\lambda +2)}
\ee
\be
E\{\mathcal{Q}^2\psi_0(\mathcal{Q})^2\}  =  \frac{ N (N+2) (\lambda + N)}{4(N +\lambda +2)}
\ee

Then, the coefficients \eqref{a1} and \eqref{a2} for the \textit{t}-distribution are:
\be
a_{1,t} \triangleq -\frac{1}{2(N +\lambda +2)}, \; a_{2,t} \triangleq \frac{ \lambda + N}{2(N +\lambda +2)}.
\ee

By substituting the previous results in \eqref{C_mu_p}, \eqref{FIM_Sigma_gk} and \eqref{cov_mat_eff_scat}, we obtain the matrices $\mb{C}_0(\mb{s}_{\bs{\mu}_0})$, $\mb{C}_0(\mb{s}_{\vsigmat})$ and $\mb{C}_0(\bar{\mb{s}}_{\vsigmat})$ for the $t$-distribution.

%

\begin{figure}[H]
	\centering
	\includegraphics[height=6cm]{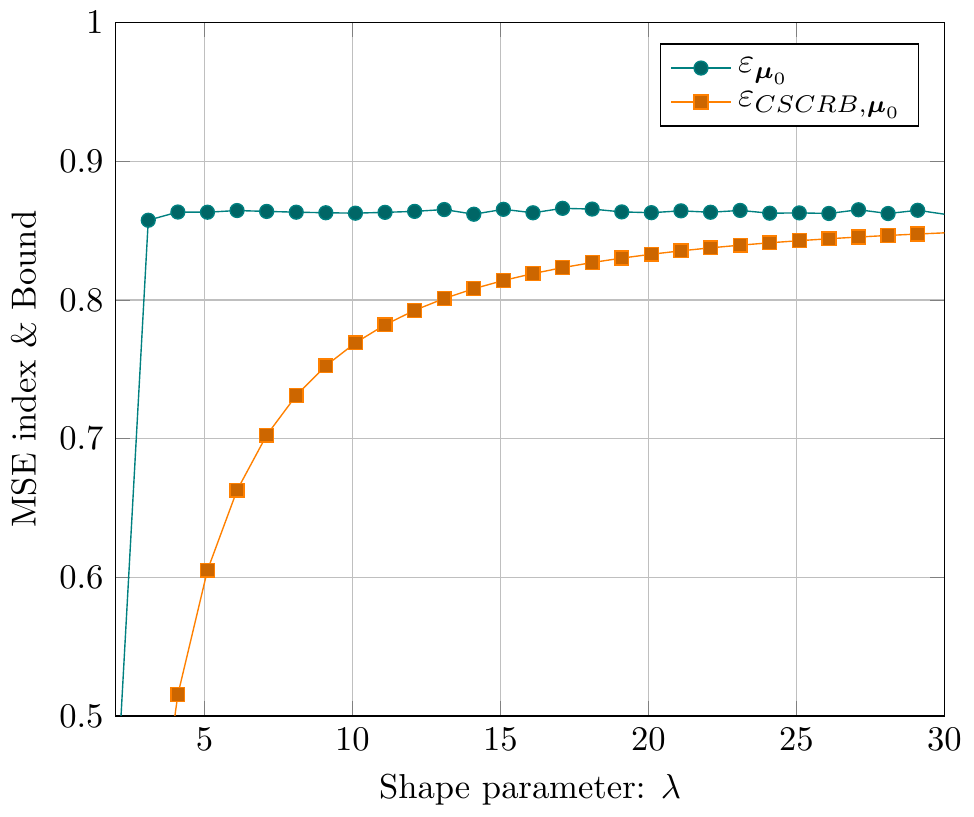}
	\caption{MSE index for $\hat{\bs{\mu}}$ and the related CSCRB as functions of the shape parameter $\lambda$ for \textit{t}-distributed data ($M=3N$).}
	\label{fig:Fig1}
\end{figure}

\begin{figure}[H]
	\centering
	\includegraphics[height=6cm]{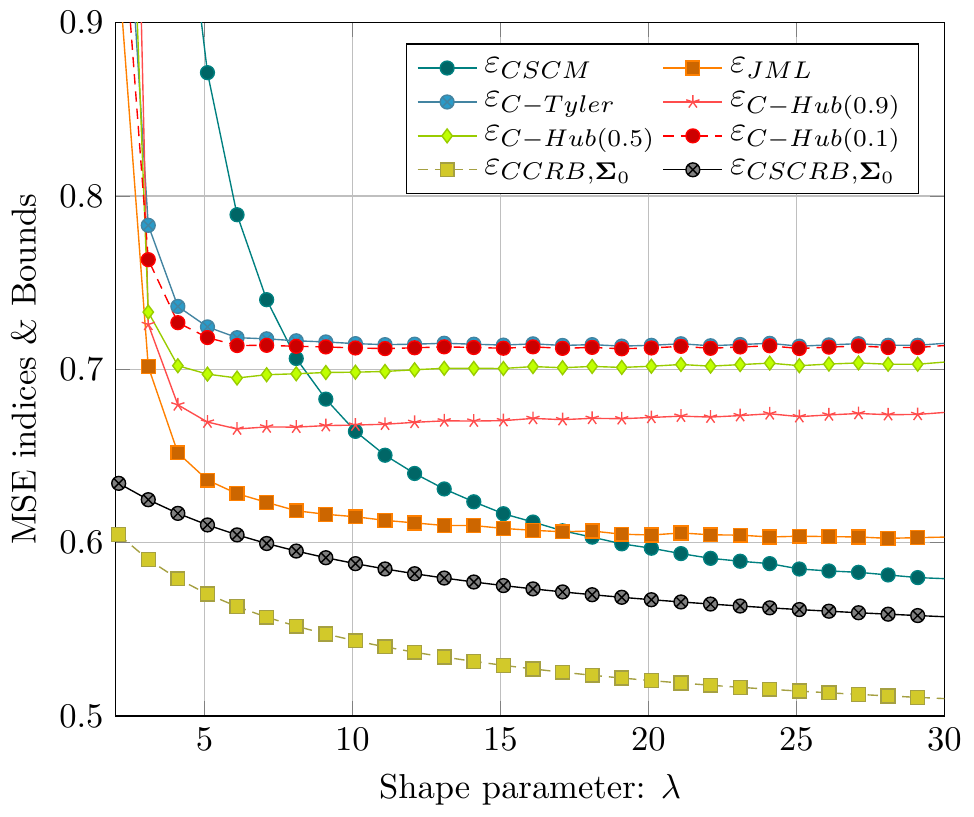}
	\caption{MSE indices for the constrained scatter matrix estimators and the related CCRB and CSCRB as functions of the shape parameter $\lambda$ for \textit{t}-distributed data ($M=3N$).}
	\label{fig:Fig2}
\end{figure}

Fig. \ref{fig:Fig1} shows the MSE index of the sample mean compared with the CSCRB as function of the shape parameter $\lambda$. As we can see, when $\lambda \rightarrow \infty$, the sample mean tends to be an efficient estimator. This is an expected result, since, when the shape parameter $\lambda$ goes to infinity, the $t$-distribution becomes a Gaussian distribution, then the sample mean is the ML estimator for $\bs{\mu}_0$. In Fig. \ref{fig:Fig2} we can observe an interesting fact: the distance between the CSCRB and the CCRB increases as $\lambda \rightarrow \infty$. This means that the lack of knowledge of the particular density generator, i.e. the lack of knowledge of the particular RES distribution of the data, has an higher impact when the tails of the true distribution become lighter. This behavior has been already observed in \cite{Hallin_P_2006}. Regarding the \textit{constrained} scatter matrix estimators, the CSCM achieves the CSCRB as $\lambda \rightarrow \infty$, i.e. as the data tends to be Gaussian distributed. Note that while it is well-known that the SCM is the ML estimator for the \textit{unconstrained} scatter matrix, the CSCM is \textit{not} the ML estimator for the constrained scatter matrix $\bs{\Sigma}_0$. This is why, as $\lambda \rightarrow \infty$, the CSCM does not achieve the CCRB. Regarding C-Tyler's and C-Huber's estimators, from Fig. \ref{fig:Fig2} we can see that C-Huber's estimator has better estimation performance for all the three values of $u$ with respect to C-Tyler's estimator. In particular, the estimation performance of C-Huber's estimator improves as $u$ tends to 1, i.e. when it tends to collapse to the CSCM. Both C-Tyler's and C-Huber's estimators are far from being efficient with respect to the CSCRB. However, it must be mentioned here that efficiency is not the only property that an estimator should have. Robustness is also important in the choice of an estimation algorithm. We will investigate the trade-off between efficiency and robustness in future works. 

Another important question that may arise is related to the behaviour of the Maximum Likelihood estimator of the scatter matrix. To answer this question, firstly we have to note that the density generator of the $t$-distribution in \eqref{g_t} depends on two additional parameters: the shape $\lambda$ and the scale $\eta$. If we assume to know perfectly \textit{both} the functional form of $g_0$ and the scale and shape parameters, then the ML estimator of the scatter matrix $\bs{\Sigma}$ will outperform all M-estimators and will achieve the classical CRB for $\bs{\Sigma}$. This scenario is discussed in \cite{Grec3}. A more realistic situation is when only the functional form of $g_0$ is assumed to be a priori known while the scale and shape parameters have to be jointly estimated with the scatter matrix $\bs{\Sigma}$. A joint ML (JML) estimator for $\bs{\Sigma}$, $\lambda$ and $\eta$ generally does not exists, so we have to rely of sub-optimal strategies as the one discussed in \cite{Fortunati2016}. Specifically, the recursive joint estimator of $\bs{\Sigma}$, $\lambda$ and $\eta$ proposed in \cite{Fortunati2016} exploits the Method of Moments (MoM) to estimate the scale and shape parameters, while an estimation of the scatter matrix is obtained using an ML approach where the unknown parameters $\lambda$ and $\eta$ are replaced by their MoM estimates. Clearly, this JML algorithm will no longer achieve the classical CRB on the scatter matrix due to the lack of a priori knowledge about the scale and shape parameters. It would be interesting to investigate how the JML estimator behaves with respect to the CSCRB. As we can see from Fig. \ref{fig:Fig1}, the MSE index of the JML is larger than the CSCRB and this would suggest that not knowing the shape and scale parameters has the same impact of not knowing the whole functional form of the density generator. Of course, this aspect deserves further investigation and we leave it to future work.  

\subsection{Case 2: the Generalized Gaussian distribution}
The density generator relative to the Generalized Gaussian (GG) distribution is:
\be
g_0(t) \triangleq \frac{2^{N/2}s\Gamma(N/2)}{\pi^{N/2}(2b)^{\frac{N}{2s}}\Gamma({N/2s)}} \exp \left( -\frac{t^s}{2b}\right)
\ee
and then $\psi_0(t)$ in \eqref{psi} is given by $\psi_0(t) = -s(2b)^{-1} t^{s-1}$.

Consequently, from \eqref{SS_Q_pdf}, we have that:
\be
p_\mathcal{Q}(q) = \frac{sq^{N/2-1}}{(2b)^{\frac{N}{2s}}\Gamma({N/2s)}}\exp\left(-\frac{q^s}{2b}\right).
\ee

Using the integral in \cite[pp. 370, n. 3.478 (1)]{Integrals}, we have that:
\be
E\{\mathcal{Q}\psi_0(\mathcal{Q})\} = -N/2.
\ee
\be
E\{\mathcal{Q}\psi_0(\mathcal{Q})^2\} = \frac{s^2\Gamma(\frac{N+4s-2}{2s})}{(2b)^{1/s}\Gamma({N/2s)}}
\ee
\be
E\{\mathcal{Q}^2\psi_0(\mathcal{Q})^2\} = N(N+2s)/4.
\ee

Then, the coefficients \eqref{a1} and \eqref{a2} for the GG distribution are:
\be
a_{1,GG} \triangleq \frac{s-1}{2(N +2)}, \; a_{2,GG} \triangleq \frac{N + 2s}{2(N +2)}.
\ee

As before, by substituting the previous results in \eqref{C_mu_p}, \eqref{FIM_Sigma_gk} and \eqref{cov_mat_eff_scat}, we obtain the matrices $\mb{C}_0(\mb{s}_{\bs{\mu}_0})$, $\mb{C}_0(\mb{s}_{\vsigmat})$ and $\mb{C}_0(\bar{\mb{s}}_{\vsigmat})$ for the GG distribution.

%

\begin{figure}[H]
	\centering
	\includegraphics[height=6cm]{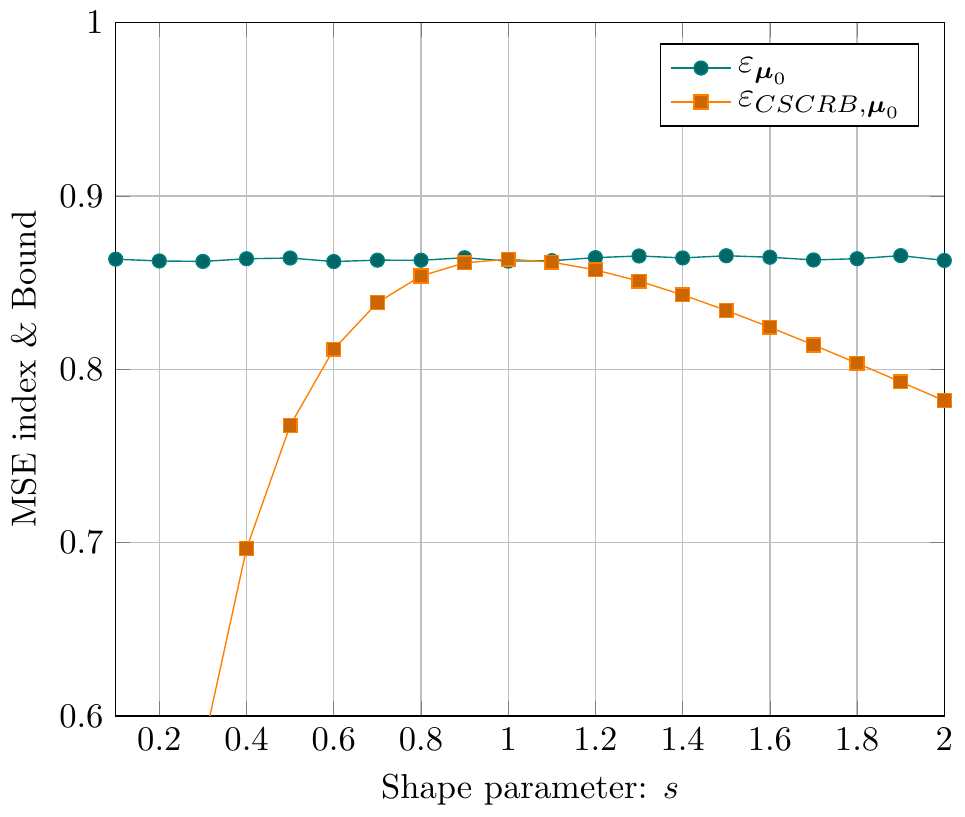}
	\caption{MSE index for $\hat{\bs{\mu}}$ and the related CSCRB as functions of the shape parameter $s$ for GG distributed data ($M=3N$).}
	\label{fig:Fig3}
\end{figure}

\begin{figure}[H]
	\centering
	\includegraphics[height=6cm]{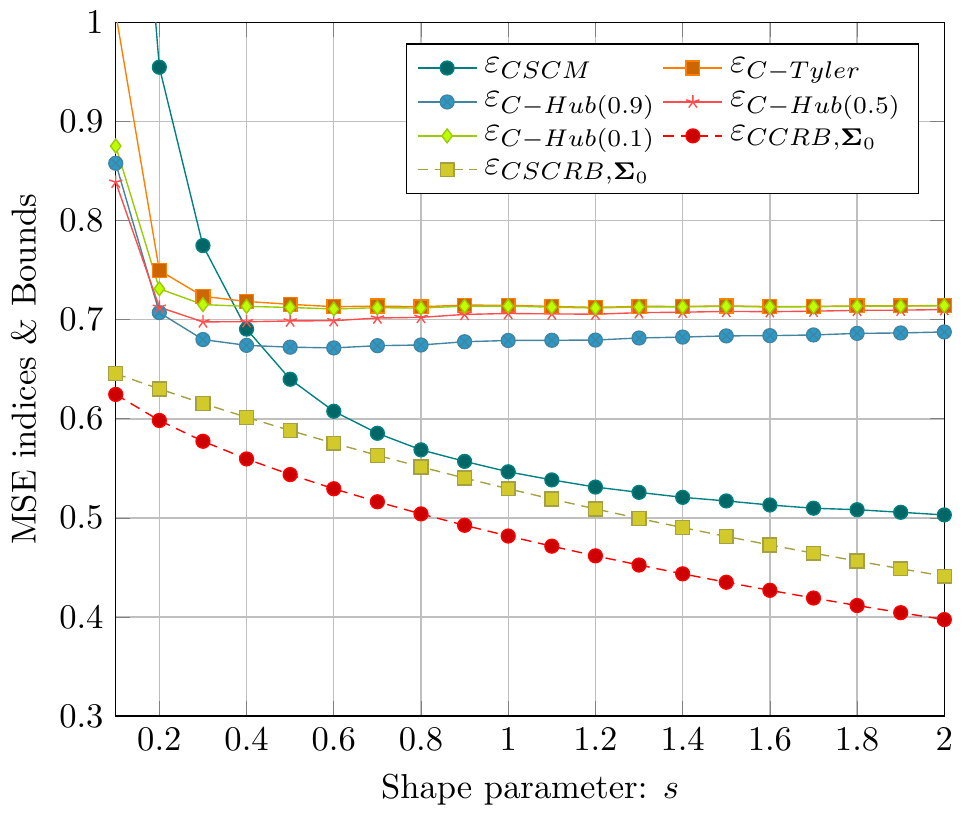}
	\caption{MSE indices for the constrained scatter matrix estimators and the related CCRB and CSCRB as functions of the shape parameter $s$ for GG distributed data ($M=3N$).}
	\label{fig:Fig4}
\end{figure}

The simulation results for the GG distributed data confirm all the previous discussions about $t$-distributed data:
\begin{itemize}
	\item The sample mean estimator is an efficient estimator of $\bs{\mu}_0$ when the data is Gaussian distributed. In fact, as we can see from Fig. \ref{fig:Fig3} that the MSE index $\varepsilon_{\bs{\mu}_0}$ equates the CSCRB for $s=1$, i.e., when the GG distribution becomes the Gaussian one.
	\item The distance between the CSCRB and the CCRB increases as the tails of the true data distribution become lighter. This behavior can be observed in Fig. \ref{fig:Fig4}. It is worth recalling here that for $0<s<1$ the GG distribution has heavier tails and for $s>1$ lighter tails compared to the Gaussian distribution that can be obtained for $s=1$.
	\item The CSCM is an efficient estimator for $\bs{\Sigma}_0$ w.r.t. the CSCRB when the data is Gaussian distributed, i.e., when $s=1$ (see Fig. \ref{fig:Fig4}). However, it does not achieve the CCRB since, as discussed before, the CSCM is not the ML estimator for the constrained scatter matrix under Gaussian distributed data.
\end{itemize}

\section{Conclusion}
\label{conclusions}
This paper is organized in two interrelated parts. The first part is devoted to place the class of RES distributions within the framework of semiparametric group models. This analysis allows to look at the well-known RES family from a different and enlightening standpoint. The main features of the semiparametric group models have been presented and discussed, paying particular attention to their implications on the family of RES distributions. In the second part of the paper, we showed how the direct application of these properties leads to derive a closed-form expresson for the CSCRB for the joint estimation of the mean vector $\bs{\mu}_0$ and of the constrained scatter matrix $\bs{\Sigma}$ of a set of RES distributed random vectors. 

Even if the semiparametric inference offers us a wide range of research opportunities, a huge amount of work still remains to be done. Our short-term research activity will be devoted to the extension of the CSCRB to the complex field, i.e. to the joint estimation of a \textit{complex} mean vector and a \textit{complex} scatter matrix in the family of Complex Elliptically Symmetric (CES) distributions. This is of great relevance in radar applications where both the data and the parameters to be estimated are modeled as complex quantities. Regarding the long-term research activities, our efforts will be devoted to an in-depth study of the robustness property of an estimator in the framework of semiparametric models. Specifically, particular attention will be devoted to the analysis of the trade-off between robustness and semiparametric efficiency of an estimator.


%

\appendices
\section{The Hilbert space of $q$-dimensional random functions}
\label{Hilber_random}
This Appendix provides some additional details on the Hilbert space $\mathcal{H}^q$ of the zero-mean, $q$-dimensional functions and on the projection operator $\Pi(\cdot|\mathcal{V})$, where $\mathcal{V} \subseteq \mathcal{H}^q$. The following discussion does not claim to provide a complete mathematical characterization of these two elements, but could be useful as background material for the derivation of the CSCRB provided in this paper. 

Let us introduce the underlying probability space $(\mathcal{X},\mathfrak{F},\mathrm{P}_X)$, where $\mathcal{X}\subseteq \mathbb{R}^N$ is the sample space, $\mathfrak{F}$ is the Borel $\sigma$-algebra of events in $\mathcal{X}$ and $\mathrm{P}_X$ is the probability measure. Let  $\mb{x} \in \mathcal{X}$ be a random vector, then $P_X(\mb{a}) \triangleq \mathrm{P}_X(x_1 \leq a_1,\ldots, x_N \leq a_N)$ is its cumulative distribution function (cdf). We assume that the cdf $P_X$ admits a relevant probability density function (pdf) (with respect to the standard Lebesgue measure), denoted as $p_X$, such that $dP_X(\mb{a})=p_X(\mb{a})d\mb{a}$.

Consider now the vector space of the one-dimensional square-integrable function defined on $(\mathcal{X},\mathfrak{F},\mathrm{P}_X)$:
\be
L_2(P_X) = \lbr h:\mathcal{X} \rightarrow \mathbb{R} \left|  \int_{\mathcal{X}} |h(\mb{x})|^2 dP_X(\mb{x}) < \infty \right.  \rbr,
\ee 
with the inner product given by:
\be
\label{H_inner_prod}
\innerprod{h}{g}_{X}=E_{X}\{hg\} \triangleq \int_{\mathcal{X}} h(\mb{x})g(\mb{x})dP_X(\mb{x}), \; \forall h,g \in L_2(P_X).
\ee

Let us define $\mathcal{H}^1 \subseteq L_2(P_X)$ as the subspace of all the one-dimensional, zero-mean functions on $(\mathcal{X},\mathfrak{F},\mathrm{P}_X)$ such that:
\be
\label{H_1_space}
\mathcal{H}^1 = \lbr h \in L_2(P_X) \left|  E_{X}\{h\} =0 \right.  \rbr.
\ee
It is immediate to verify that $\mathcal{H}^1$, endowed with the inner product defined in \eqref{H_inner_prod}, is an infinite-dimensional Hilbert space \cite[Ch. 2]{Tsiatis}.

The \virg{$q$-replicating} Hilbert space $\mathcal{H}^q$ of the zero-mean, $q$-dimensional functions on $(\mathcal{X},\mathfrak{F},\mathrm{P}_X)$ is defined as the Cartesian product of $q$ copies of the Hilbert space $\mathcal{H}^1$, i.e. $\mathcal{H}^q \triangleq \mathcal{H}^1 \times \mathcal{H}^1 \times \cdots \times \mathcal{H}^1$ \cite[Ch. 3, Def. 6]{Tsiatis}, such that:
\be
\label{H_q_space}
\mathcal{H}^q = \lbr \mb{h}=[h_1,\ldots,h_q]^T \left| h_i \in \mathcal{H}^1,\; i=1,\ldots,q \right.  \rbr.
\ee

Due to the Cartesian product-based construction, the inner product of $\mathcal{H}^q$ is naturally induced by the one of $\mathcal{H}^1$, i.e.:
\be
\label{inner_in_H}
\innerprod{\mb{h}_1}{\mb{h}_2}_{X} \triangleq E_{X}\{\mb{h}^T_1\mb{h}_2\} = \sum\nolimits_{i=1}^{q} E_{X}\{h_{1,i}h_{2,i}\}, 
\ee
and consequently the norm is:
\be
\label{norm_in_H}
\norm{\mb{h}}_{X}=\sqrt{\innerprod{\mb{h}}{\mb{h}}_{X}} = \sum\nolimits_{i=1}^{q} E_{X}\{h_i^2\}.
\ee 

Let us now investigate the geometrical structure of $\mathcal{H}^q$, with a particular focus on the orthogonal projection of a generic element $\mb{h} \in \mathcal{H}^q$ into a closed subspace $\mathcal{V}$ of $\mathcal{H}^q$. The following theorem is a fundamental result in Hilbert spaces theory, and can be established in a very general setting (see e.g. \cite[Theo. 3.9.3]{Hilbert}). Here, we will adapt it to the particular Hilbert space $\mathcal{H}^q$.

\begin{theorem}[The Projection Theorem]
	\label{proj_theo}
	Let $\mathcal{V}$ be a closed subspace of the Hilbert space $\mathcal{H}^q$ and let $\mb{h}$ and $\mb{g}$ be two $q$-dimensional, zero-mean functions on $(\mathcal{X},\mathfrak{F},\mathrm{P}_X)$, such that $\mb{h} \in \mathcal{H}^q$ and $\mb{g} \in \mathcal{V}$ . Then, the following conditions are equivalent:
	\begin{enumerate}
		\item $\norm{\mb{h}-\mb{g}}_{X} = \inf \limits_{\mb{l} \in \mathcal{V}} \norm{\mb{h}-\mb{l}}_{X}$,
		\item $\mb{h}$ can be uniquely written as
		\be
		\label{orth_rep}
		\mb{h}=\mb{g}+(\mb{h}-\mb{g}),
		\ee
		where $\mb{g} \triangleq \Pi(\mb{h}|\mathcal{V}) \in \mathcal{V}$ and $\mb{h} - \Pi(\mb{h}|\mathcal{V}) \in \mathcal{V}^{\perp}$, where $\mathcal{V}^{\perp}$ indicates the orthogonal complement of $\mathcal{V}$ in $\mathcal{H}^q$,
		\item the element $\mb{g} \triangleq \Pi(\mb{h}|\mathcal{V}) \in \mathcal{V}$ is then uniquely determined by the orthogonality constraint
		\be
		\label{ortho_con}
		\innerprod{\mb{h}-\Pi(\mb{h}|\mathcal{V})}{\mb{l}}_{X}=0,\quad \forall \mb{l} \in \mathcal{V}.
		\ee 
	\end{enumerate}
\end{theorem}
The operator $\Pi(\cdot|\mathcal{V})$ defined in Theorem \ref{proj_theo} is called the \textit{orthogonal projection operator} onto the closed subspace $\mathcal{V}$. The unique element $\Pi(\mb{h}|\mathcal{V})$ is then called the \textit{orthogonal projection} of $\mb{h} \in \mathcal{H}^q$ onto $\mathcal{V}$. Furthermore, as a consequence of the Condition 2) in Theorem \ref{proj_theo}, the Hilbert space $\mathcal{H}^q$ can be written as the Cartesian product of the subspace $\mathcal{V}$ and of its orthogonal complement $\mathcal{V}^{\perp}$, i.e. $\mathcal{H}^q = \mathcal{V} \times \mathcal{V}^{\perp}$.

\section{Projection operator and conditional expectation}   
\label{cond_expt_sect}
In Appendix \ref{Hilber_random} we defined $\mathcal{H}^q$ as the Hilbert space of the $q$-dimensional, zero-mean functions on the probability space $(\mathcal{X},\mathfrak{F},\mathrm{P}_X)$. Let $\mathfrak{G}(V) \subseteq \mathfrak{F}$ be the sub-sigma algebra generated by the random variable $V$. It can be shown (see e.g. \cite[Ch. 23]{Jacod} and \cite[Appendix 3]{BKRW}) that the set of all the $q$-dimensional, zero-mean functions on the probability space $(\mathcal{X},\mathfrak{G}(V),\mathrm{P}_{X|V})$ is a closed linear subsapce, say $\mathcal{V}$, of the Hilbert space $\mathcal{H}^q$. 

This fact can be exploited to establish a link between the projection operator $\Pi(\cdot|\mathcal{V})$ and the \textit{conditional expectation} $E_{X|V}\{\cdot|V\}$. Firstly, let us define $E_{X|V}\{\cdot|V\}$ as in \cite[Ch. 23]{Jacod} and \cite[Appendix 3]{BKRW}.

\begin{definition}
	\label{cond_expt}
	Let $\mb{h} \in \mathcal{H}^q$ and $\mb{l} \in \mathcal{V} \subseteq \mathcal{H}^q$ be two $q$-dimensional, zero-mean  functions on the probability spaces $(\mathcal{X},\mathfrak{F},\mathrm{P}_X)$ and $(\mathcal{X},\mathfrak{G}(V),\mathrm{P}_{X|V})$ with $\mathfrak{G}(V) \subseteq \mathfrak{F}$, respectively. Then the conditional expectation $E_{X|V}\{\mb{h}|V\}$ is the unique element in $\mathcal{V}$, such that:
	\be
	\label{cond_exp_prop}
	\innerprod{\mb{h}-E_{X|V}\{\mb{h}|V\}}{\mb{l}}_{X} \triangleq E_{X}\{(\mb{h}-E_{X|V}\{\mb{h}|V\})^T\mb{l}\}= 0,
	\ee
	for every $\mb{l} \in \mathcal{V}$.
\end{definition}
For a more general and formal definition we refer the reader to \cite[Ch. 23]{Jacod} and \cite[Appendix 3]{BKRW}.
The condition \eqref{cond_exp_prop} is equivalent to \eqref{ortho_con} in Theorem \ref{proj_theo} that defines the projection operator, and consequently, we have that:
\be
\Pi(\cdot|\mathcal{V})=E_{X|V}\{\cdot|V\}.
\ee

The usefulness of this relation is in the fact that, for some semiparametric models, the tangent space presents an invariance structure with respect to a group of transformations and it admits a characterization through a certain sub-sigma algebra generated by the relevant maximal invariant statistic \cite{Lehmann_TSH}. An example of a semiparametric model that owns this property is the semiparametric group model of RES distributions discussed in Subsection \ref{RES_section_group_mod}. 

\section*{Acknowledgment}
The work of Stefano Fortunati has been partially supported by the Air Force Office of Scientific Research under award number FA9550-17-1-0065.




%

\bibliographystyle{IEEEtran}
\bibliography{ref1}

\end{document}